\documentclass[12pt,aps,prb]{revtex4}
\usepackage{epsfig}
\usepackage{amsmath}

\normalsize

\def\b0{{\bf 0}}

\def\cO{{\cal O}}

\def\tG{\tilde G}

\def\Re{{\rm Re}}
\def\Im{{\rm Im}}

\def\lra{\leftrightarrow}
\def\alf{\alpha}
\def\eps{\epsilon}

\def\Gam{\Gamma}
\def\lam{\lambda}
\def\Lam{\Lambda}
\def\om{\omega}
\def\Om{\Omega}

\def\Sg{\Sigma}

\def\lra{\leftrightarrow}

\begin{document}


\title{\large Functional renormalization group for Luttinger liquids 
 \\ with impurities}

\author{S. Andergassen,$^1$ T. Enss,$^1$ V. Meden,$^2$ 
 W. Metzner,$^1$ \\
 U. Schollw\"ock,$^3$ and K. Sch\"onhammer$^2$ \\
 {\small\em $^1$Max-Planck-Institut f\"ur Festk\"orperforschung, 
 D-70569 Stuttgart, Germany} \\
 {\small\em $^2$Institut f\"ur Theoretische Physik, Universit\"at 
 G\"ottingen, D-37077 G\"ottingen, Germany} \\
 {\small\em $^3$Institut f\"ur Theoretische Physik, Technische Hochschule
 Aachen, D-52056 Aachen, Germany}}

\date{\small March 19, 2004}



\begin{abstract}
We improve the recently developed functional renormalization group
(fRG) for impurities and boundaries in Luttinger liquids by 
including renormalization of the two-particle interaction, in addition 
to renormalization of the impurity potential.
Explicit flow equations are derived for spinless lattice fermions
with nearest neighbor interaction at zero temperature, and a fast 
algorithm for solving these equations for very large systems is 
presented.
We compute spectral properties of single-particle excitations,
and the oscillations in the density profile induced by impurities
or boundaries for chains with up to $10^6$ lattice sites. 
The expected asymptotic power-laws at low energy or long distance
are fully captured by the fRG. Results on the relevant energy scales 
and crossover phenomena at intermediate scales are also obtained.
A comparison with numerical density matrix renormalization results
for systems with up to 1000 sites shows that the fRG with the
inclusion of vertex renormalization is remarkably accurate even
for intermediate interaction strengths. 

\noindent
\mbox{PACS: 71.10.Pm, 72.10.-d, 73.21.Hb} \\
\end{abstract}

\maketitle


\section{Introduction}

Low dimensional electron systems exhibit a rich variety of surprising 
effects which are due to the cooperative interplay of impurities
and interactions. 
In one dimension even clean metallic systems are always strongly 
affected by interactions: at low energy scales physical properties
obey anomalous power-laws, known as Luttinger liquid behavior, 
which is very different from the conventional Fermi liquid behavior 
describing most higher dimensional metals.\cite{Gia,Voi}
In Luttinger liquids with repulsive interactions already
a single static impurity is known to affect the low-energy properties 
drastically.\cite{LP,Mat,AR,GS,KF,EA,YGM}
The impurity potential in a repulsive Luttinger liquid becomes
dressed by long-range oscillations which suppress the spectral
weight for single-particle excitations near the impurity and also the 
conductance through the impurity down to zero in the low-energy limit.

The {\em asymptotic} low-energy properties of Luttinger liquids with a
single impurity are rather well understood. Universal power-laws
and scaling functions have been obtained by bosonization, conformal
field theory, and exact solutions for the low-energy asymptotics in 
special integrable cases.\cite{Gia10}
What remains to be developed, however, is a many-body method for
{\em microscopic} models of interacting fermions with impurities,
which does not only capture correctly the universal low-energy
asymptotics, but allows one to compute observables on all energy
scales, providing thus also {\em non-universal} properties, and in
particular an answer to the important question at what {\em scale} the
ultimate asymptotics sets in. That scale can indeed be surprisingly
low, and the properties above it very different from the asymptotic
behavior.
Some of the non-universal properties can be computed numerically 
by the density matrix renormalization group (DMRG),\cite{Sch} 
but this method is limited to lattice systems with about 1000 sites, 
and only a restricted set of observables can be evaluated with 
affordable computational effort.

In the last few years it has been realized that the functional 
renormalization group (fRG) is a source of powerful new computation 
tools for interacting Fermi systems, especially for low-dimensional 
systems with competing instabilities and entangled infrared 
singularities.
The starting point of this approach is an exact hierarchy of 
differential flow equations for the Green or vertex functions of
the system, which is obtained by taking derivatives with respect to
an infrared cutoff $\Lam$.\cite{Sal}
Approximations are then constructed by truncating the hierarchy and 
parametrizing the vertex functions with a manageable set of variables 
or functions.\cite{2dHM}

A relatively simple fRG approximation for impurities in spinless 
Luttinger liquids has been developed recently by some of 
us.\cite{MMSS}
The scheme starts from an fRG hierarchy for one-particle irreducible
vertex functions, as first derived in a field theoretical context 
by Wetterich\cite{Wet} and Morris,\cite{Mor} and for interacting Fermi 
systems by Salmhofer and Honerkamp.\cite{SH}
A Matsubara frequency cutoff is used as the flow parameter.
The hierarchy is then truncated already at first order, such that the
2-particle vertex remains unrenormalized, and the flow of the
self-energy, which describes the renormalized impurity potential, 
is determined by the bare 2-particle interaction. 
No simplified parametrization of the self-energy is necessary such 
that the full spatial dependence of the renormalized impurity potential 
can be obtained for very large lattice systems.
In spite of the striking simplicity of this scheme it was shown that
the effects of a single static impurity in a spinless Luttinger liquid
are fully captured qualitatively, and in the weak coupling limit also
quantitatively.\cite{MMSS}
In particular, one obtains that impurity potentials in repulsive 
Luttinger liquids become effectively stronger at lower energy scales,
and act ultimately as a weak link between two otherwise separate
wires, as predicted by Kane and Fisher.\cite{KF}
The fRG also correctly yields the universal low energy power-laws with 
exponents that do not depend on the bare impurity strength.
In addition, it was shown that the asymptotic behavior holds typically 
only at very low energy scales and for very large systems, except for
very strong bare impurities.\cite{MMSS}
The fRG approach to impurities in Luttinger liquids was originally
developed and tested for spectral densities of single-particle
excitations, but has been applied very recently also to transport 
problems, such as persistent currents in mesoscopic rings\cite{MS1,MS2}
and the conductance of interacting wires connected to non-interacting
leads.\cite{MS2,MAX} 
The full power of the fRG with its ability to deal naturally also 
with complex crossover phenomena emerges most convincingly in 
the multi-scale problem posed by the transport through a resonant 
double barrier.\cite{MEX}

In the present work we further develop the fRG approach for impurities
in Luttinger liquids by including 2-particle vertex renormalization,
that is we go one step further in the hierarchy of flow equations
than previously.\cite{fn1} 
For spinless fermions this extension does not matter qualitatively,
as the lowest order is already qualitatively correct, but the
quantitative accuracy of the results improves considerably in
particular at intermediate interaction strengths.
For spin-$\frac{1}{2}$ systems, which we will treat in a subsequent
work,\cite{AEX} vertex renormalization is necessary to take into 
account that backscattering of particles with opposite spins at 
opposite Fermi points scales to zero in the low energy limit.
A crucial point is to devise an efficient parametrization of the 
vertex by a managable number of variables.
Here we focus on spinless lattice fermions with nearest neighbor
interaction as a prototype model.
The bulk model is supplemented by site or hopping impurities.
We also analyze the influence of boundaries, which can be viewed
as infinite barriers or infinitesimal weak links.
We choose to parametrize the vertex by a renormalized 
nearest neighbor interaction, which allows us to capture various
advantageous features:
the low energy flow of the vertex at $k_F$ in the pure system is 
obtained correctly to second order in the renormalized couplings;
the non-universal contributions at finite energy scales are correct
to second order in the bare interaction;
the algorithm for the flow of the self-energy remains as fast as
in the absence of vertex renormalization, such that one can easily 
deal with up to $10^6$ lattice sites.
We compute spectral properties of single-particle
excitations near an impurity and the oscillations in the density 
profile induced by an impurity, or by a boundary.
The accuracy of the calculation is checked by comparing with
DMRG results for systems with up to 1000 lattice sites and with
exact results for the asymptotic behavior, which can be obtained
from the Bethe ansatz and bosonization.

The article is structured as follows.
In Sec.\ II we introduce the microscopic model and various types
of impurities. 
The fRG formalism is developed in Sec.\ III, and worked out
explicitly for the spinless fermion model with nearest neighbor 
interaction.
Sec.\ IV is dedicated to results for the renormalized impurity
potential, spectral properties, and the density profile.
We finally conclude in Sec.\ V with an outline of promising 
extensions of the present work.

\section{Microscopic model}

We consider the spinless fermion lattice model with nearest
neighbor interaction and various types of impurity potentials.
The Hamiltonian has the form
\begin{equation}
 H = H_0 + H_I + H_{\rm imp}
\end{equation}
where the kinetic energy
\begin{equation}
 H_0 = -t \sum_j \big( \,
 c^{\dag}_{j+1} c_j + c^{\dag}_j \, c_{j+1} \, \big)
\end{equation}
is given by nearest neighbor hopping with an amplitude $t$, and
\begin{equation}
 H_I = U \sum_j n_j \, n_{j+1}
\end{equation}
is a nearest neighbor interaction of strength $U$.
We use standard second quantization notation, where $c^{\dag}_j$ 
and $c_j$ are creation and annihilation operators on site $j$,
respectively, and $n_j = c^{\dag}_j \, c_j$ is the local density
operator.
The impurity is represented by
\begin{equation}
 H_{\rm imp} = \sum_{j,j'} 
 V_{j'j} \; c^{\dag}_{j'} \, c_j
\end{equation}
where $V_{j'j}$ is a static potential.
For ''site impurities'' 
\begin{equation}
 V_{j'j} = V_j \, \delta_{jj'}
\end{equation}
this potential is local. For the important special case of a
single site impurity
\begin{equation}
 V_j = V \, \delta_{jj_0}
\end{equation}
the potential acts only on one site $j_0$.
We also consider ''hopping impurities'' described by the non-local
potential 
\begin{equation}
 V_{j'j} = V_{jj'} = - t_{j,j+1} \, \delta_{j',j+1} \; .
\end{equation}
For the special case of a single hopping impurity,
\begin{equation}
 t_{j,j+1} = (t'-t) \, \delta_{jj_0} \; ,
\end{equation}
the hopping amplitude $t$ is replaced by $t'$ on the bond linking
the sites $j_0$ and $j_0+1$.
In the following we will set the bulk hopping amplitude $t$ equal
to one, that is all energies are given in units of $t$.

The clean spinless fermion model $H_0 + H_I$ is exactly solvable
via the Bethe ansatz.\cite{YY}
The system is a Luttinger liquid for all particle densities $n$ 
and any interaction strength except at half-filling 
for $|U| > 2$. For $U>2$ a charge density wave
with wave vector $\pi$ forms, for $U<-2$ the system undergoes
phase separation.
The Luttinger liquid parameter $K_{\rho}$, which determines all
the critical exponents of the liquid, can be computed exactly
from the Bethe ansatz solution.\cite{Hal}
At half-filling $K_{\rho}$ is related to $U$ by the simple 
explicit formula
\begin{equation}
 K_{\rho}^{-1} = \frac{2}{\pi} \, 
 \arccos \left(-\frac{U}{2} \right)
\end{equation}
for $|U| \leq 2$.

We will compute physical properties for the above models on
finite chains with open boundary conditions. 
The fRG equations can be solved easily for chains with up to 
$10^6$ sites, the DMRG can be applied for up to $10^3$ sites. 
To avoid oscillations emerging from the ends of the chain we
will sometimes attach non-interacting semi-infinite leads to
the finite interacting chain, with a smoothly decaying interaction 
at the contact.

\section{Functional renormalization group}

This section is dedicated to a detailed presentation of the
fRG equations for one-dimens\-ional Fermi systems with impurities, 
including vertex renormalization, at zero temperature.
We use the one-particle irreducible version of the functional
RG.\cite{Wet,Mor}
The starting point is an exact hierarchy of flow equations for
the irreducible vertex functions, which is obtained by
cutting off the infrared part of the free propagator on a scale
$\Lam$, and differentiating the generating functional for the 
vertex functions with respect to this scale.\cite{SH}

\subsection{Cutoff and flow equations}

The cutoff can be imposed in many different ways.
The only requirement is that the infrared singularities must be 
regularized such that the flow equations allow for a regular 
perturbation expansion in powers of the renormalized two-particle
vertex.
For our purposes a sharp Matsubara frequency cutoff is the most 
efficient choice. A momentum cutoff is less suitable here since
the impurity spoils momentum conservation.
The cutoff is imposed by excluding modes with frequencies below 
scale $\Lam$ from the functional integral representation of the
system, or equivalently, by introducing a regularized bare 
propagator
\begin{equation}
 G_0^{\Lam}(i\om) = \Theta(|\om| - \Lam) \, G_0(i\om) \; .
\end{equation}
Here $G_0$ is the bare propagator of the pure system, involving
neither interactions nor impurities.
Instead of the sharp cutoff imposed by the step function $\Theta$
one may also choose a smooth cutoff function, but the sharp cutoff 
has the advantage that it reduces the number of integration variables
on the right hand side of the flow equations. 
Note that we will frequently write expressions which are well defined
only if the sharp cutoff is viewed as a limit of increasingly sharp
smooth cutoff functions.
The suppression of frequencies below scale $\Lam$ affects all Green 
and vertex functions of the interacting system, which become thus 
functions of $\Lam$. The original system is recovered in the limit 
$\Lam \to 0$.

The first two equations in the exact hierarchy of flow equations for 
the one-particle irreducible $m$-particle vertex functions are 
depicted graphically in Figs.\ 1 and 2. 
The first equation yields the flow of the self-energy, which is related
to the interacting propagator by the usual Dyson equation
\begin{equation}
 G^{\Lam} = \left[ (G_0^{\Lam})^{-1} - \Sg^{\Lam} \right]^{-1}
 \; .
\end{equation}
Here and below $G^{\Lam}$, $\Sg^{\Lam}$ etc.\ are operators, which
do not refer to any particular single-particle basis, unless we
write matrix indices explicitly. 
The right hand side of the flow equation involves the 2-particle
vertex $\Gam^{\Lam}$, and the so-called single scale propagator
$S^{\Lam}$ defined as
\begin{equation}
 S^{\Lam} \, = \,
 G^{\Lam} \left[ \partial_{\Lam} (G_0^{\Lam})^{-1} \right] 
 G^{\Lam} \, = \, - \,
 \frac{1}{1 - G_0^{\Lam} \, \Sg^{\Lam}} \; 
 \frac{\partial G_0^{\Lam}}{\partial\Lam} \;
 \frac{1}{1 - \Sg^{\Lam} \, G_0^{\Lam}} \; .
\end{equation}
Note that $S^{\Lam}$ has support only on a single frequency
scale: $|\om| = \Lam$. 
The precise flow equation for the self-energy (sketched graphically 
in Fig.\ 1) reads
\begin{equation}
 \frac{\partial}{\partial\Lam} \Sg^{\Lam}(1',1) =
 \, - \, \frac{1}{\beta} \, \sum_{2,2'} \, e^{i\om_2 0^+} \,
 S^{\Lam}(2,2') \; \Gam^{\Lam}(1',2';1,2) \; ,
\end{equation}
where $\beta$ is the inverse temperature. The numbers 1, 2, etc.\
are a shorthand for Matsubara frequencies and labels for 
single-particle states such as site and spin indices. 
Note that $\om_1 = \om'_1$ and $\om_2 = \om'_2$ due to 
Matsubara frequency conservation.
The exponential factor in the above equation is irrelevant at
any finite $\Lam$, but is necessary to define the initial
conditions of the flow at $\Lam = \Lam_0 \to \infty$ (see
below).

The right hand side of the flow equation for the 2-particle 
vertex $\Gam^{\Lam}$, shown in Fig.\ 2, involves $\Gam^{\Lam}$
itself, but also the 3-particle vertex $\Gam_3^{\Lam}$.
The latter could be computed from a flow equation involving
the 4-particle vertex $\Gam_4^{\Lam}$, and so on.
To avoid this unmanagable proliferation of vertex functions we 
make our first approximation:
we neglect the contribution of the 3-particle vertex to the flow 
of $\Gam^{\Lam}$.
The coupled system of flow equations for $\Gam^{\Lam}$ and 
$\Sg^{\Lam}$ is then closed. 
The contribution of $\Gam_3^{\Lam}$ to $\Gam^{\Lam}$ is small
as long as $\Gam^{\Lam}$ is sufficiently small, because 
$\Gam_3^{\Lam}$ is initially (at $\Lam_0$) zero and is generated 
only from terms of third order in $\Gam^{\Lam}$.\cite{SH}
A comparison of the RG results to exact DMRG results and exact 
scaling properties shows that the truncation error is often
surprisingly small even for rather large interactions.\cite{MMSS}
The explicit form of the truncated flow equation for the 2-particle 
vertex reads
\begin{eqnarray}
 \frac{\partial}{\partial\Lam} \Gam^{\Lam}(1',2';1,2)
 &=& \; \frac{1}{\beta} \,
 \sum_{3,3'} \sum_{4,4'} \, G^{\Lam}(3,3') \, S^{\Lam}(4,4')
 \nonumber \\[2mm]
 \times & \Big[
 & \! \Gam^{\Lam}(1',2';3,4) \, \Gam^{\Lam}(3',4';1,2) 
 \nonumber \\[2mm]
 &-& \!  \Gam^{\Lam}(1',4';1,3) \, \Gam^{\Lam}(3',2';4,2) 
 - (3 \lra 4, 3' \lra 4') \nonumber \\[2mm]
 &+& \! \Gam^{\Lam}(2',4';1,3) \, \Gam^{\Lam}(3',1';4,2)
 + (3 \lra 4, 3' \lra 4') 
 \, \Big]
\end{eqnarray}
The various contributions are shown diagrammatically in Fig.\ 3.

For a sharp frequency cutoff the frequency sums on the right
hand side of the flow equations can be carried out analytically
in the zero temperature limit, where the Matsubara sums become 
integrals.
At this point one has to deal with products of delta functions 
$\delta(|\om| - \Lam)$ and expressions involving step functions 
$\Theta(|\om| - \Lam)$.
These at first sight ambiguous expressions are well defined and
unique if the sharp cutoff is implemented as a limit of increasingly 
sharp broadened cutoff functions $\Theta_{\eps}$, with the broadening 
parameter $\eps$ tending to zero. 
The expressions can then be conveniently evaluated by using the 
following relation,\cite{Mor} valid for arbitrary continuous
functions $f$:
\begin{equation}
 \delta_{\eps}(x-\Lam) \, f[\Theta_{\eps}(x-\Lam)] \to
 \delta(x-\Lam) \int_0^1 f(t) \, dt \; ,
\end{equation}
where $\delta_{\eps} = \Theta'_{\eps}$. 
Note that the functional form of $\Theta_{\eps}$ for finite $\eps$ 
does not affect the result in the limit $\eps \to 0$.

Instead of writing down the frequency-integrated flow equations in
full generality, we first implement our second approximation:
the frequency dependent flow of the renormalized 2-particle vertex 
$\Gam^{\Lam}$ is replaced by its value at vanishing (external) 
frequencies, such that $\Gam^{\Lam}$ remains frequency independent.
As a consequence, also the self-energy becomes frequency independent.
Since the bare interaction is frequency independent, neglecting the
frequency dependence leads to errors only at second order (in the 
interaction strength) for the self-energy, and at third order for
the vertex function at zero frequency.
Besides the quantitative errors we miss qualitative properties
related to the frequency dependence of the self-energy, such as the
suppression of the one-particle spectral weight in the bulk of
a pure Luttinger liquid.
On the other hand, a comparison with exact numerical results and
asymptotic analytical results shows that the impurity effects are
not qualitatively affected by the frequency dependence of $\Sg$, 
at least for weak interactions.

Carrying out the frequency integration in the flow equation for the
self-energy, and inserting a frequency independent 2-particle vertex,
one obtains
\begin{equation}
 \frac{\partial}{\partial\Lam} \Sg^{\Lam}_{1',1} =
 - \frac{1}{2\pi} \sum_{\om = \pm \Lam} \sum_{2,2'} \,
 e^{i\om 0^+} \, \tG^{\Lam}_{2,2'}(i\om) \,
 \Gam^{\Lam}_{1',2';1,2} \; ,
\end{equation}
where the lower indices $1$, $2$, etc.\ label single-particle states 
(not frequencies) and
\begin{equation}
 \tG^{\Lam}(i\om) = 
 \left[ G_0^{-1}(i\om) - \Sg^{\Lam} \right]^{-1} \; .
\end{equation}
Note that $\tG^{\Lam}$ has no jump at $|\om| = \Lam$, in contrast
to $G^{\Lam}$.
The frequency integrated flow equation for the 2-particle vertex, 
evaluated at vanishing external frequencies, has the form
\begin{eqnarray}
 \frac{\partial}{\partial\Lam} \Gam^{\Lam}_{1',2';1,2} \, =
 \hskip 1cm && \nonumber \\[2mm] 
 \frac{1}{2\pi} \, 
 \sum_{\om = \pm\Lam} \, \sum_{3,3'} \sum_{4,4'} \!
 & \Big[ & \!\! \frac{1}{2} \,
 \tG^{\Lam}_{3,3'}(i\om) \, \tG^{\Lam}_{4,4'}(-i\om) \,
 \Gam^{\Lam}_{1',2';3,4} \, \Gam^{\Lam}_{3',4';1,2} 
 \nonumber \\[2mm]
 & + & \!\! \tG^{\Lam}_{3,3'}(i\om) \, \tG^{\Lam}_{4,4'}(i\om) \,
 \left( - \Gam^{\Lam}_{1',4';1,3} \, \Gam^{\Lam}_{3',2';4,2} 
        + \Gam^{\Lam}_{2',4';1,3} \, \Gam^{\Lam}_{3',1';4,2}
 \right) \Big]
\end{eqnarray}

The flow is determined uniquely by the differential flow equations 
and the initial conditions at $\Lam = \infty$. 
For $\Lam = \infty$ the vertex functions are given by the bare
interactions of the system. In particular, the flow of the 2-particle 
vertex starts from the antisymmetrized bare 2-particle interaction 
while $m$-particle vertices of higher order vanish at $\Lam = \infty$,
in the absence of $m$-body interactions with $m>2$.
The self-energy at $\Lam = \infty$ is given by the bare one-particle 
potential, that is by those one-particle terms which are not
included already in $G_0$. We choose to include the translationally
invariant bulk kinetic energy and the chemical potential in $G_0$,
while the impurity (site or hopping) potential $V$ is assigned to the
initial condition of the self-energy.
In a numerical solution the flow starts at some large finite
initial cutoff $\Lam_0$. Here one has to take into account that, 
due to the slow decay of the right hand side of the flow equation
for $\Sg^{\Lam}$ at large $\Lam$, the integration of the flow from 
$\Lam = \infty$ to $\Lam = \Lam_0$ yields a contribution which does
not vanish in the limit $\Lam_0 \to \infty$, but rather tends to a
finite constant.
Since $\tG^{\Lam}_{2,2'}(i\om) \to \delta_{2,2'}/(i\om)$ for
$|\om| = \Lam \to \infty$, this constant is easily determined as
\begin{equation}
 - \frac{1}{2\pi} \lim_{\Lam_0 \to \infty} \int_{\infty}^{\Lam_0}
 d\Lam \sum_{\om = \pm\Lam} \sum_{2,2'} e^{i\om 0^+} \,
 \frac{\delta_{2,2'}}{i\om} \, I_{1',2';1,2} =
 \frac{1}{2} \sum_{2} I_{1',2;1,2} \; ,
\end{equation}
where $I_{1',2';1,2}$ is the bare antisymmetrized interaction.
In summary, the initial conditions for the self-energy and the
2-particle vertex at $\Lam = \Lam_0 \to \infty$ are
\begin{eqnarray}
 \Sg^{\Lam_0}_{1,1'} &=& 
 V_{1,1'} + \frac{1}{2} \sum_{2} I_{1',2;1,2} \\[2mm]
 \Gam^{\Lam_0}_{1',2';1,2} &=& I_{1',2';1,2} \; ,
\end{eqnarray}
where $V_{1,1'}$ is the bare impurity potential.
For the flow at $\Lam < \Lam_0$ the factor $e^{i\om 0^+}$ in 
Eq.\ (16) can be discarded.

\subsection{Parametrization of 2-particle vertex}

For a finite lattice system with $L$ sites the flow of the 2-particle
vertex $\Gam^{\Lam}_{1',2';1,2}$ involves $\cO(L^3)$ independent 
flowing variables, if translation invariance is assumed, and 
$\cO(L^4)$ variables, if the influence of the impurity on the flow 
of the 2-particle vertex is taken into account.
For a treatment of large systems it is therefore necessary to reduce
the number of variables by a suitable approximate parametrization of
the vertex. 
In the low energy limit (small $\Lam$) the flow is dominated by a
very small number of variables anyway, the others being irrelevant
according to standard RG arguments.\cite{Voi}
In particular, the frequency dependence of the vertex, discarded 
already above, is irrelevant for the flow of $\Gam^{\Lam}$ at 
small $\Lam$.
For larger $\Lam$ one can use perturbation theory as a guide for a
simple but efficient parametrization of $\Gam^{\Lam}$.

We neglect the feedback of the self-energy into the
flow of $\Gam^{\Lam}$ and replace $\tG^{\Lam}$ by $G_0$ in Eq.\ (18).
The renormalization of $\Gam^{\Lam}$ includes thus only bulk, not
impurity, contributions, such that $\Gam^{\Lam}$ remains translation
invariant. 
While this turns out to be sufficient for capturing the effects of
isolated impurities in otherwise pure systems, it is known that
impurity contributions to vertex renormalization become important
in macroscopically disordered systems.\cite{GM}
For a more concrete treatment of the vertex renormalization, we have 
to focus on a specific model. In this article we restrict ourselves
to the spinless fermion model with nearest neighbor interaction,
leaving an extension to spin-$\frac{1}{2}$ systems with more
general interactions for subsequent work.\cite{AEX}

For spinless fermions the 2-particle vertex and the self-energy are
fully characterized by either site or momentum variables.
In the low energy limit, the flow of the vertex is dominated by
contributions with momenta close to the Fermi points, such that the
right hand side of the flow equation is determined by momentum 
components of the vertex $\Gam^{\Lam}_{k'_1,k'_2;k_1,k_2}$ with 
$k_1,k_2,k'_1,k'_2 = \pm k_F$. 
Due to the antisymmetry of the vertex, there is only one such 
component which is non-zero:
\begin{equation}
 g^{\Lam} = \Gam^{\Lam}_{k_F,-k_F;k_F,-k_F} \; .
\end{equation}
In the low energy limit the momentum dependence of the vertex away
from $\pm k_F$ is irrelevant. There are therefore many possible 
choices for the functional form of $\Gam^{\Lam}_{k'_1,k'_2;k_1,k_2}$,
which all lead to the correct low energy asymptotics. 
For a model with a bare nearest neighbor interaction $U$, a natural
and efficient choice is to parametrize the flowing vertex simply by
a renormalized nearest neighbor interaction $U^{\Lam}$, which leads
to a real space vertex of the form 
\begin{equation}
 \Gam^{\Lam}_{j'_1,j'_2;j_1,j_2} =
 U^{\Lam}_{j_1,j_2} \, 
 ( \delta_{j_1,j'_1} \delta_{j_2,j'_2} - 
   \delta_{j_1,j'_2} \delta_{j_2,j'_1} )
\end{equation}
with $U^{\Lam}_{j_1,j_2} = U^{\Lam} \, 
(\delta_{j_1,j_2-1} + \delta_{j_1,j_2+1}) \,$.
This yields the following structure in momentum space:
\begin{equation}
 \Gam^{\Lam}_{k'_1,k'_2;k_1,k_2} =
 2 U^{\Lam} \, [ \cos(k'_1 - k_1) - \cos(k'_2 - k_1) ] \,
 \delta^{(2\pi)}_{k_1+k_2,k'_1+k'_2} \; ,
\end{equation}
where the Kronecker delta implements momentum conservation 
(modulo $2\pi$).
The flowing coupling constant $U^{\Lam}$ is linked to the value of
the vertex at the Fermi points by the relation
\begin{equation}
 g^{\Lam} = 2 U^{\Lam} \, [ 1 - \cos(2k_F) ] \; .
\end{equation}
The flow equation for $g^{\Lam}$ becomes
\begin{equation}
 \frac{\partial g^{\Lam}}{\partial\Lam} =
 \frac{1}{2\pi} \sum_{\om = \pm\Lam} \int \frac{dp}{2\pi} \,
 (\, {\rm PP + PH + PH'} \,)
\end{equation}
with the particle-particle and particle-hole contributions
\begin{eqnarray}
 {\rm PP} &=& \frac{1}{2} \, G^0_p(i\om) \, G^0_{-p}(-i\om) \,
 \Gam^{\Lam}_{k_F,-k_F;p,-p} \, \Gam^{\Lam}_{p,-p;k_F,-k_F} 
 \nonumber \\[2mm]
 {\rm PH} &=& - [G^0_p(i\om)]^2 \,
 \Gam^{\Lam}_{k_F,p;k_F,p} \, \Gam^{\Lam}_{p,-k_F;p,-k_F} 
 \nonumber \\[2mm]
 {\rm PH'} &=& G^0_{p-k_F}(i\om) \, G^0_{p+k_F}(i\om) \,
 \Gam^{\Lam}_{-k_F,p+k_F;k_F,p-k_F} \, 
 \Gam^{\Lam}_{p-k_F,k_F;p+k_F,-k_F} \; ,
\end{eqnarray}
where $\Gam^{\Lam}$ on the right hand side of the flow equation
is given by Eq.\ (24). 
Using Eq.\ (25) to replace $\partial_{\Lam} g^{\Lam}$ by 
$\partial_{\Lam} U^{\Lam}$ on the left hand side of Eq.\ (26),
one obtains a flow equation for $U^{\Lam}$ of the simple form
\begin{equation}
 \partial_{\Lam} U^{\Lam} = h(\Lam) \, (U^{\Lam})^2 \; .
\end{equation}
The function $h(\Lam)$ depends only on the cutoff $\Lam$ and the
Fermi momentum $k_F$. An explicit formula for $h(\Lam)$ can be 
obtained by carrying out the momentum integral in Eq.\ (26) using 
the residue theorem, as sketched in Appendix A.
For finite systems the momentum integral should be replaced by a
discrete momentum sum; however, this leads to sizable corrections
only for very small systems.
The flow equation (28) can be integrated to
\begin{equation}
 U^{\Lam} = \frac{U}{1 - U \, H(\Lam)} \; ,
\end{equation}
where $H(\Lam)$ is the primitive function of $h(\Lam)$ with
$H(\Lam) \to 0$ for $\Lam \to \infty$.
Integrating $h(\Lam)$ one obtains
\begin{eqnarray}
 H(\Lam) &=& - \frac{\Lam}{2\pi} + \frac{1}{\pi} \, \Re \Big[ \,
 \frac{(4-\mu_0^2)\Lam^2 - 2i\mu_0(2-\mu_0^2)\Lam + \mu_0^4 - 6\mu_0^2 
 + 8}{2 \, (4-\mu_0^2)\sqrt{\Lam^2 - 2i\mu_0\Lam + 4 - \mu_0^2}}
 - \frac{i\mu_0}{2} \, \sinh^{-1} \frac{\Lam-i\mu_0}{2} 
 \nonumber \\[2mm]
 && + \, \frac{\mu_0^4}{2(4-\mu_0^2)^{3/2}} \, \tanh^{-1}
 \frac{4 + \mu_0^2 + i\mu_0\Lam}
 {\sqrt{(4 + \mu_0^2 + i\mu_0\Lam)^2 + 4(\Lam-2i\mu_0)^2}} \Big] \; ,
\end{eqnarray}
where $\mu_0 = -2\cos k_F$, while $\sinh^{-1}$ and $\tanh^{-1}$
denote the main branch of the inverse of the complex functions 
$\sinh$ and $\tanh$, respectively.
At half-filling, corresponding to $k_F = \pi/2$, the function 
$h(\Lam)$ is particularly simple (see Appendix A), such that also
$U^{\Lam}$ has a relatively simple form:
\begin{equation}
 U^{\Lam} = \frac{U}
 {1 + 
 \left(\Lam - \frac{2 + \Lam^2}{\sqrt{4 + \Lam^2}} \right) \, 
 U/(2\pi)} \; .
\end{equation}
In Fig.\ 4 we show results for the renormalized nearest neighbor
interaction $U^{\Lam}$ as obtained from the flow equation at various
densities $n$, for a bare interaction $U = 1$. 
While the renormalization does not follow any simple rule
at intermediate scales $\Lam$, all curves saturate at a finite value 
$U^*$ in the limit $\Lam \to 0$, corresponding to a finite $g^*$,
as expected for a Luttinger liquid fixed point.\cite{Voi}

The above simplified flow equation yields not only the correct 
low energy asymptotics to second order in the renormalized vertex,
but contains also all {\em non-universal}\/ second order corrections 
to the vertex at $\pm k_F$ at higher energy scales. 
Hence the resulting fixed point coupling $g^*$, from which the Luttinger 
liquid parameter $K_{\rho}$ can be computed, is obtained correctly 
to second order in $U$. The calculation of $K_{\rho}$, presented in 
Appendix B, illustrates how Luttinger liquid parameters can be related 
to the fixed point couplings obtained from the fRG. 
A comparison of the RG result for $K_{\rho}$ with exact results from 
the Bethe ansatz solution of the spinless fermion model shows that the 
vertex renormalization scheme described above is not only very simple, 
but also surprisingly accurate.

Parametrizing $\Gam^{\Lam}$ by a renormalized nearest neighbor 
interaction has the enormous advantage that the self-energy, 
as determined by the flow equation (16), is a tridiagonal matrix 
in real space, that is only the matrix elements $\Sg^{\Lam}_{j,j}$
and $\Sg^{\Lam}_{j,j \pm 1}$ are non-zero.
Inserting $\Gam^{\Lam}$ from Eq.\ (23) into (16), one obtains the
following simple coupled flow equations for the diagonal and 
off-diagonal matrix elements: 
\begin{eqnarray}
 \frac{\partial}{\partial\Lam} \, \Sg^{\Lam}_{j,j} \; =&
 - & \frac{U^{\Lam}}{2\pi} \sum_{\om = \pm\Lam} \sum_{r = \pm 1}
 \, \tG^{\Lam}_{j+r,j+r}(i\om) \nonumber \\[2mm]
 \frac{\partial}{\partial\Lam} \, \Sg^{\Lam}_{j,j \pm 1} \; =&
   & \frac{U^{\Lam}}{2\pi} \sum_{\om = \pm\Lam}
 \tG^{\Lam}_{j,j \pm 1}(i\om) \; .
\end{eqnarray}
Note that the self-energy enters also the right hand side of these
equations, via $\tG^{\Lam} = (G_0^{-1} - \Sg^{\Lam})^{-1}$.
Since $\Sg^{\Lam}$ and $G_0^{-1}$ are both tridiagonal in real
space, the matrix inversion required to compute the diagonal and
first off-diagonal elements of $\tG^{\Lam}$ from $\Sg^{\Lam}$
can be carried out very efficiently, and thus for very large 
systems. In Appendix C we describe an algorithm for the numerical 
solution of the flow equation for $\Sg$ which scales only linearly 
with the size of the system.

\subsection{Observables}

In the next section we will present results for spectral 
properties of single-particle excitations near an impurity
or boundary, and for the density profile. Here we describe how 
the relevant observables are computed from the solution of
the flow equations.

\subsubsection{Single-particle excitations}

Integrating the flow equation for the self-energy $\Sg^{\Lam}$
down to $\Lam = 0$ yields the physical (cutoff-free) self-energy
$\Sg$, and the single-particle propagator 
$G = (G_0^{-1} - \Sg)^{-1}$.
From the latter spectral properties of single-particle 
excitations can be extracted.
We focus on local spectral properties, which are described by 
the \emph{local spectral function}
\begin{equation}
 \rho_j(\om) = - \frac{1}{\pi} \, 
 \Im \, G_{jj}(\om + i0^+) \; ,
\end{equation}
where $G_{jj}(\om + i0^+)$ is the local propagator, analytically
continued to the real frequency axis from above.

In our approximation the self-energy is frequency-independent
and can therefore be viewed as an effective single-particle
potential. The propagator $G$ is thus the Green function of an 
effective single-particle Hamiltonian. 
In real space representation this Hamiltonian is given by the
tridiagonal matrix $h_{\rm eff} = h_0 + \Sg$, where the matrix 
elements of $h_0$ are the hopping amplitudes in $H_0$, Eq.\ (2).
For a lattice with $L$ sites this matrix has $L$ (including
possible multiplicities) eigenvalues $\eps_{\lam}$ and an 
orthonormal set of corresponding eigenvectors $\psi_{\lam}$.
For the spectral function $\rho_j(\om)$ one thus obtains a 
sum of $\delta$-functions
\begin{equation}
 \rho_j(\om) = \sum_{\lam} w_{\lam j} \, 
 \delta(\om - \xi_{\lam}) \; ,
\end{equation}
where $\xi_{\lam} = \eps_{\lam} - \mu$, and the \emph{spectral weight} 
$w_{\lam j}$ is the squared modulus of the amplitude of $\psi_{\lam}$ 
on site $j$. For large $L$ the level spacing between neighboring
eigenvalues is usually of order $L^{-1}$, except for one or a few 
levels outside the band edges which correspond to bound states.

Due to even-odd effects etc.\ the spectral weight $w_{\lam j}$ 
generally varies quickly from one eigenvalue to the next one. 
A smooth function of $\om$ which suppresses these usually
irrelevant finite-size details can be obtained by averaging
over neighboring eigenvalues. In addition it is useful to divide
by the level spacing between eigenvalues to obtain the 
conventional local \emph{density of states}, which we denote by 
$D_j(\om)$.

\subsubsection{Density profile}

Boundaries or impurities induce a density profile with long-range 
Friedel oscillations, which are expected to decay with a power-law
at long distances.\cite{EG}
The expectation value of the local density $n_j$ could be
computed from the local one-particle propagator $G_{jj}$, if
$G$ was known exactly. However, the approximate flow equations
for $\Sg$ can be expected to describe the asymptotic behavior
of $G$ correctly only at long distances between creation and
annihilation operator in time and/or space, while in the local 
density operator time and space variables coincide. In the
standard RG terminology $n_j$ is a \emph{composite} operator,
which has to be renormalized separately.\cite{Zin}

To derive a flow equation for $n_j$, we follow the usual procedure
for the renormalization of correlation functions involving
composite operators: we add a term $\phi_j n_j$ with a small
field $\phi_j$ to the Hamiltonian, and take derivatives with
respect to $\phi_j$ in the flow equations. 
The local density is given by 
\begin{equation}
 n_j = \left. \frac{\partial\Om(\phi_j)}{\partial\phi_j}  
 \right|_{\phi_j = 0} \; ,
\end{equation}
where $\Om(\phi_j)$ is the grand canonical potential of the 
system in the presence of the field $\phi_j$.
Note that we use the same symbol $n_j$ for the density operator 
and its expectation value.
In the presence of a cutoff $\Lam$ the grand canonical potential
obeys the exact flow-equation
\begin{equation}
 \frac{\partial}{\partial\Lam} \Om^{\Lam} = 
  \frac{1}{\beta} \, 
  \sum_{\om} \, {\rm tr} \left\{ e^{i\om 0^+} \, 
  [\partial_{\Lam} G_0^{\Lam}(i\om)^{-1}] \, 
  [G^{\Lam}(i\om) - G_0^{\Lam}(i\om)]
  \right\}  
\end{equation}
in a short-hand matrix notation. This flow-equation follows
from the functional flow equation for the generating functional 
for vertex functions \cite{SH} and the relation between the 
grand canonical potential and the zero-particle vertex.
At zero temperature the Matsubara frequency sum becomes an 
integral which, for the sharp frequency cutoff (10), 
can be carried out analytically. This yields
\begin{equation}
 \frac{\partial}{\partial\Lam} \Om^{\Lam} = 
 \frac{1}{2\pi}
 \sum_{\om = \pm\Lam} {\rm tr} \left\{ e^{i\om 0^+}
 \log[ 1 - G_0(i\om) \, \Sg^{\Lam}(i\om) ] \right\} \; .
\end{equation}
We attribute $\phi_j$ to the ''interaction'' part of the 
Hamiltonian, not to $H_0$, such that $G_0$ remains independent
of $\phi_j$. 
The self-energy is modified via the additional local and 
frequency-independent contribution $\phi_j \, \delta_{jj'}$
to its initial value $\Sg^{\Lam_0}_{jj'}$ at scale $\Lam_0$.

The density profile can be obtained from the above equations
and the flow equation for $\Sg^{\Lam}$ by computing the shift
of $\Om$ generated by a small finite perturbation $\phi_j$,
that is by numerical differentiation.
Alternatively one may carry out the $\phi_j$-derivative
analytically in the flow equations, which yields a flow equation 
for the density in terms of the density response vertex.
Taking the $\phi_j$ derivative in (37) yields
\begin{equation}
 \frac{\partial}{\partial\Lam} n_j^{\Lam} = 
 - \frac{1}{2\pi} \sum_{\om = \pm\Lam} 
 {\rm tr} \left[ e^{i\om 0^+}
 \tG^{\Lam}(i\om) \, R_j^{\Lam}(i\om) \right]
\end{equation}
with the density response vertex
\begin{equation}
 R_j^{\Lam}(i\om) = 
 \left. \frac{\partial\Sg^{\Lam}(i\om)}{\partial\phi_j}
 \right|_{\phi_j = 0} \; ,
\end{equation}
where the propagator $\tG^{\Lam}$ is computed as previously,
that is in the absence of $\phi_j$.
We compute the self-energy $\Sg^{\Lam}$ in the presence of $\phi_j$ 
within the same approximation as previously. It is thus determined
from the flow equation (16) with a frequency-independent 
two-particle vertex $\Gam^{\Lam}$. Taking a derivative of that
equation with respect to $\phi_j$ at $\phi_j = 0$ yields the flow 
equation for the response vertex
\begin{equation}
 \frac{\partial}{\partial\Lam} R_{j;1',1}^{\Lam} = 
 - \frac{1}{2\pi} \sum_{\om = \pm\Lam} 
 \sum_{2,2'} \sum_{3,3'} 
 \tG_{2,3'}^{\Lam}(i\om) \, R_{j;3',3}^{\Lam} \, 
 \tG_{3,2'}^{\Lam}(i\om) \,
 \Gam_{1',2';1,2}^{\Lam} \; .
\end{equation}
Note that $R_j^{\Lam}$ is frequency-independent in our 
approximation and that there is no contribution from the 
$\phi_j$-derivative of $\Gam^{\Lam}$, since we neglect 
self-energy contributions in the flow of the two-particle 
vertex.

For spinless fermions with a (renormalized) nearest neighbor
interaction, the matrix $R_j^{\Lam}$ is tridiagonal, that is
only the components $R_{j;l,l}^{\Lam}$ and 
$R_{j;l,l\pm 1}^{\Lam}$ are non-zero, and their flow is given
by
\begin{eqnarray}
 \frac{\partial}{\partial\Lam} R_{j;l,l}^{\Lam} &=&
 - \frac{U^{\Lam}}{2\pi} \sum_{\om=\pm\Lam}
 \sum_{l'} \sum_{r=\pm 1} \sum_{r' = 0,\pm 1}
 \tG_{l+r,l'}^{\Lam}(i\om) \, R_{j;l',l'+r'}^{\Lam} \,
 \tG_{l'+r',l+r}^{\Lam}(i\om) \nonumber \\[2mm]
 \frac{\partial}{\partial\Lam} R_{j;l,l\pm 1}^{\Lam} &=&
 \frac{U^{\Lam}}{2\pi} \sum_{\om=\pm\Lam}
 \sum_{l'} \sum_{r' = 0,\pm 1}
 \tG_{l,l'}^{\Lam}(i\om) \, R_{j;l',l'+r'}^{\Lam} \,
 \tG_{l'+r',l\pm 1}^{\Lam}(i\om) \; .
\end{eqnarray} 
The right hand sides involve only one unrestricted
lattice summation, and can be computed in ${\cal O}(L)$ time
(see Appendix C). The initial condition for the response vertex 
is $R_{j;l,l'}^{\Lam_0} = \delta_{jl} \delta_{ll'}$.
The initial condition for the density is
$n_j^{\Lam_0} = \frac{1}{2}$, for any filling, due to the slow 
convergence of the flow equation (38) at large frequencies, which 
yields a finite contribution to the integrated flow from $\Lam = 
\infty$ to $\Lam_0$ for arbitrarily large finite $\Lam_0$,
as in the case of the self-energy discussed in more detail
above.

To avoid the interference of Friedel oscillations emerging
from the impurity or one boundary with those coming from the 
(other) boundaries of our systems we suppress the influence 
of the latter by coupling the finite chain to semi-infinite
non-interacting leads, with a smooth decay of the interaction
at the contacts.
The presence of leads modifies the self-energy in the 
interacting region only via a boundary term. In particular,
a lead coupled to the first site of the interacting region 
via nearest neighbor hopping (with amplitude $t=1$) yields 
the additional contribution
\begin{equation}
 \Sg_{j,j'}^{\rm lead}(i\om) = 
 \frac{i\om+\mu_0}{2} \left( 1 - 
 \sqrt{1 - \frac{4}{(i\om+\mu_0)^2}} \, \right) 
 \, \delta_{1j} \delta_{jj'}
\end{equation}
to the first diagonal element of self-energy matrix in real 
space.\cite{cond} 
Note that this term is frequency dependent and independent of
$\Lam$. A lead coupled to the last site yields an analogous
contribution to the last diagonal element of the self-energy.

\section{Results}

We now present and discuss explicit results for the self-energy, 
spectral properties near the impurity or boundary, 
and the density profile, as obtained from the fRG. 
A comparison with exact DMRG results is made for the spectral
weight at the Fermi level and for the density profile.
Some fRG results for the self-energy and spectral 
properties of the spinless fermion model have been published 
already previously,\cite{MMSS} 
but only for half-filling and without any vertex renormalization. 
Here we present results also away from half-filling and 
demonstrate the quantitative improvement of accuracy obtained by 
including vertex renormalization. 
In addition we present for the first time fRG results 
for the density profile.
Using a faster algorithm (see Appendix C) than previously we are
now able to solve the flow equations for much larger systems.

The typical shape of the self-energy in the vicinity of an impurity 
can be seen in Fig.\ 5, where we plot the diagonal elements 
$\Sg_{j,j}$ and the off-diagonal elements $\Sg_{j,j+1}$ near a
site impurity of strength $V=1.5$ added to the spinless fermion
model with interaction strength $U=1$ at quarter-filling. 
Recall that the self-energy is tridiagonal in real
space and frequency independent within our treatment.
The diagonal elements can be interpreted as a local effective 
potential, the off-diagonal elements as a non-local effective 
potential which renormalizes the hopping amplitudes.
At long distances from the impurity both $\Sg_{j,j}$ and
$\Sg_{j,j+1}$ tend to a constant. The former describes just a bulk
shift of the chemical potential, the latter a bulk renormalization
of the hopping amplitude toward larger values.
The oscillations around the bulk shifts are generated by the
impurity. The wave number of the oscillations is $2k_F=\pi/2$, 
where $k_F$ is the Fermi wave vector of the bulk system at 
quarter-filling. 
The amplitude of the oscillations of $\Sg$ decays slower than
the inverse distance from the impurity at intermediate length
scales, but approaches a decay proportional to $1/|j-j_0|$ for
$|j-j_0| \to \infty$.
This can be seen most clearly by plotting an effective exponent
$\beta_j$ for the decay of the oscillations, defined as the
negative logarithmic derivative of the oscillation amplitude 
with respect to the distance $|j-j_0|$.
In Fig.\ 6 we show the effective exponent resulting from the 
oscillations of $\Sg_{j,j}$ as a function of the distance from 
a site impurity, for $U=1$ and half-filling. 
The impurity is situated at the center of a long chain with 
$L = 2^{18}+1$ sites. 
To avoid interferences with oscillations from the boundaries we 
have attached semi-infinite non-interacting leads to the ends 
of the interacting chain, as described in the last paragraph of 
III.C.2. 
Only for relatively large impurity strengths the asymptotic regime 
corresponding to $\beta_j = 1$ is reached before finite size 
effects set in. For small $V$ one can see that $\beta_j$ increases 
from values below one, but the asymptotic long-distance behavior 
is cut off by the boundaries of the interacting region. For very small $V$
(for example $V=0.01$ in Fig.\ 6) we observe a plateau in $\beta_j$ for
intermediate distances from the impurity site. In this regime $\beta_j$
is close to $K_\rho$ which can be understood by analytically solving our flow 
equations for small $V$.\cite{MMSS}

The long-range $2k_F$-oscillations of the self-energy lead to
a marked suppression of the spectral weight for single-particle
excitations at the Fermi level, that is at $\om = 0$.
In Fig.\ 7 we show the local density of states $D_j(\om)$ on
the site next to a site impurity of strength $V=1.5$ for the
spinless fermion model at half-filling. The result for the 
interacting system at $U=1$ is compared to the non-interacting 
case. Even-odd effects have been eliminated by averaging over 
neighboring eigenvalues (see Sec.\ III.C.1). 
$\delta$-peaks outside the band edges corresponding to bound states 
are not plotted. 
The interaction leads to a global broadening of the band, which
is due to an enhancement of the bulk hopping amplitude, and also
to a strong suppression of $D_j(\om)$ at low frequencies which
is not present in the non-interacting system. 
For a finite system (here $L=1025$) the spectral weight at the
Fermi level remains finite, but tends to zero with increasing 
system size.
In Fig.\ 8 we show results for the density of states choosing 
the same parameters as in Fig.\ 7, but now for densities away
from half-filling: $n=1/4$ and $n=3/4$. 
In addition to the dip near $\om=0$ 
a second singularity appears at a finite frequency. This effect
is due to the fact that a long-range potential with a wave number 
$2k_F$ does not only strongly scatter states with momenta near 
$k_F$, but also those with momenta close to $\pi-k_F$. Indeed
the singularity is situated at $\om = \eps_{\pi-k_F} - \mu$,
where $\eps_k$ is the renormalized (bulk) dispersion.
In the half-filled case only one singularity is seen simply
because $\pi-k_F = k_F$ for $k_F = \pi/2$.

The spectral weight at the Fermi level is expected to vanish
asymptotically as a power-law $L^{-\alf_B}$ with increasing 
system size, where $\alf_B = K_{\rho}^{-1} - 1$ is the 
\emph{boundary exponent} describing the
power-law suppression of the density of states at the boundary
of a semi-infinite chain.\cite{KF} That exponent depends only
on the bulk parameters of the model, not on the impurity
strength. For the spinless fermion model it can be computed 
exactly from the Bethe ansatz solution.\cite{Hal}
We therefore analyze the asymptotic behavior of the spectral
weight at the Fermi level by defining an effective exponent 
$\alf(L)$ as the negative logarithmic derivative of the spectral 
weight with respect to the system size, such that $\alf(L)$ 
tends to a (positive) constant in case of a power-law suppression.
In Fig.\ 9 we show results for $\alf(L)$ as obtained from the
fRG for the spinless fermion model at quarter-filling 
with up to about $10^6$ sites, for a weak ($U=0.5$) and an
intermediate ($U=1.5$) interaction parameter. 
The spectral weight has been computed either at a boundary, or 
near a hopping impurity of strength $t'=0.5$. Results obtained
from the RG without (upper panel) and with (lower panel) vertex
renormalization are compared to exact DMRG results (for up to 
512 sites) and the exact boundary exponents $\alf_B$, plotted as 
horizontal lines. 
The fRG results follow a power-law for large
$L$, with the same asymptotic exponent for the boundary and
impurity case, confirming thus the expected universality.
However, the asymptotic regime is reached only for fairly large
systems, even for the intermediate interaction strength $U=1.5$.
The comparison with the exact DMRG results and exact exponents
shows that the fRG is also quantitatively rather
accurate, and that the inclusion of vertex renormalization leads
to a substantial improvement at intermediate coupling strength.
Results for the effective exponent $\alf$ in the case of a site
impurity are shown in Fig.\ 10, at quarter-filling and for an
interaction strength $U=1$. The comparison of the different
curves obtained for different impurity strengths confirms once 
again the expected asymptotic universality, and also how the 
asymptotic regime shifts rapidly toward larger systems as the 
bare impurity strength decreases.

We finally discuss results for the density profile $n_j$.
Boundaries and impurities induce Friedel oscillations of the
local density with a wave vector $2k_F$.
In a non-interacting system these oscillations decay 
proportionally to the inverse distance from the boundary or
impurity. 
In an interacting Luttinger liquid the Friedel oscillations are 
expected to decay as $|j-j_0|^{-K_{\rho}}$ at long distances 
$|j-j_0|$.
For a very weak impurity one expects a slower decay proportional
to $|j-j_0|^{1-2K_{\rho}}$ at intermediate distances, and a
crossover to the asymptotic power-law with exponent $K_{\rho}$
at very long distances.\cite{EG} At intermediate distances the
response of the density to a weak impurity can be treated in
linear response theory, such that the density modulation is
determined by the density-density response function at $2k_F$, 
which leads to the power-law decay with exponent $2K_{\rho}-1$.
In Fig.\ 11 we show fRG and DMRG results for the 
density profile $n_j$ for a spinless fermion chain with 128
sites and interaction strength $U=1$ at half-filling. 
The Friedel oscillations emerge from both boundaries and 
interfere in the center of the chain. 
The accuracy of the RG results is excellent for all $j$.
For incommensurate filling factors the density profile looks
more complicated. This can be seen in Fig.\ 12, where functional
RG results are shown for the density modulation $|n_j - n|$ 
near the boundary of a system with an average density $n=0.393$ 
and $8192$ sites. For long distances from the boundary the 
oscillation amplitude has a well-defined envelope which fits
to a power-law as a function of $j$. 
We now analyze the long-distance behavior of the amplitudes
more closely for the half-filled case, and compare to exact
results for the asymptotic exponents.
In Fig.\ 13 we show fRG results for the amplitude of
density oscillations emerging from an open boundary, for a
very long spinless fermion chain with $2^{19}+1$ sites and
various interaction strengths $U$ at half-filling.
The other end of the chain (opposite to the open boundary) is
smoothly connected to a non-interacting lead. 
In a log-log plot (upper panel of Fig.\ 13) the amplitude follows 
a straight line for almost all $j$, corresponding to a power-law 
dependence. 
Deviations from a perfect power-law can be seen more neatly by
plotting the effective exponent $\alf_j$, defined as the
negative logarithmic derivative of the amplitude with respect
to $j$ (see the lower panel of Fig.\ 13). The effective 
exponent is almost constant except at very short distances or
when $j$ approaches the opposite end of the interacting chain,
which is not surprising. From a comparison with the exact 
exponent (horizontal lines in the figure) one can assess the 
quantitative accuracy of the fRG results. 
Effective exponents describing the decay of Friedel oscillations 
generated by site impurities of various strengths are shown in 
Fig.\ 14, for a half-filled spinless fermion chain with $2^{18}+1$ 
sites and interaction $U=1$. Both ends of the interacting chains 
are coupled to non-interacting leads to suppress oscillations 
coming from the boundaries.
For strong impurities the results are close to the boundary result
(cf.\ Fig.\ 13), as expected. 
For weaker impurities the oscillations decay more slowly, that is
with a smaller exponent, and approach the boundary behavior only 
asymptotically at large distances (beyond the range of our chain
for $V<1$). For very weak impurities ($V=0.01$ in Fig.\ 14) the
oscillation amplitude follows a power-law corresponding to the 
linear response behavior with exponent $2K_{\rho} - 1$ at 
intermediate distances.
We finally present some results for the effective exponent of the
density oscillation decay in the case of an \emph{attractive} 
interaction $U=-1$, see Fig.\ 15. In that case the effective
impurity strength should scale to zero at low energies and long
distances.\cite{KF}
Indeed, for weak and moderate bare impurity potentials
the effective exponent in Fig.\ 15 approaches the linear density 
response exponent $2K_{\rho} - 1$.
Only for very strong impurities the density oscillations
decay with the smaller boundary exponent over several length scales.

\section{Conclusion}

In summary, we have shown that the fRG provides an ideal tool
for computing the intriguing properties of Luttinger liquids 
with static impurities. The method yields ''ab initio'' results
for microscopic model systems at all energy scales from the
Fermi energy to the ultimate low energy limit.
We have demonstrated the power of the method by computing
spectral properties of single-particle excitations as well as 
the oscillations in the density profile induced by impurities 
or boundaries for a spinless fermion model with nearest neighbor 
interaction. 
With the inclusion of vertex renormalization, in addition to 
the renormalization of the effective impurity potential, our
results agree remarkably well with exact asymptotic results and 
numerical DMRG data even for intermediate interaction strength.

There is a broad range of interesting further applications and 
extensions of the fRG for impurities in Luttinger liquids beyond
the scope of the present article:

{\em Spin-$\frac{1}{2}$ fermions}: The inclusion of spin degrees
of freedom requires a parametrization of the two-particle vertex
with several coupling constants. 
We have already derived flow equations for the (extended) Hubbard 
model, with an effective vertex parametrized by local and nearest 
neighbor interactions.\cite{AEX}

{\em Feedback of impurities on vertex}: For isolated impurities the
influence of impurities on the vertex renormalization is 
irrelevant for the asymptotic low energy or long distance 
behavior, although it may contribute quantitatively at intermediate
scales. In the case of disordered systems with a finite impurity
density the influence of the latter on the two-particle 
vertex is crucial and must be taken into account.\cite{GM}
In principle this is possible by computing the vertex flow
with full propagators, which contain the renormalized impurity 
potential via the self-energy.

{\em Finite temperature}: The fRG approach can be extended without
major complications to finite temperature. This is particularly
useful for studying the temperature dependence of transport
properties.\cite{cond}

{\em Transport}: In the absence of inelastic processes 
($\Im\Sg = 0$) the conductance of the interacting wire can be 
computed from the one-particle Green function in the presence
of leads.\cite{cond} Several results have already been presented
in short articles.\cite{MS2,MAX,MEX} 

{\em Inelastic processes}: Inelastic processes appear at second
order in the interaction and can be included in the flow 
equations by inserting the second order vertex into the flow
equation for the self-energy without neglecting its frequency
dependence. 
This procedure would also capture the anomalous dimension of
the bulk system, which is missing in the present work.

\vskip 1cm

\noindent
{\bf Acknowledgements:} \\
We are grateful to Manfred Salmhofer for useful discussions.
This work benefitted from the workshop ''Realistic theories of 
correlated electron materials'' in fall 2002 at ITP-UCSB.
V.M.\ thanks the Bundesministerium f\"ur Bildung und Forschung
for financial support. 

\begin{appendix}

\section{Evaluation of vertex flow}

Here we sketch some details concerning the explicit evaluation of
the flow equations for the 2-particle vertex.
Inserting the momentum structure of $\Gam^{\Lam}$, (24), 
into the flow equation (26) and replacing $g^{\Lam}$ by $U^{\Lam}$ 
on the left hand side yields
\begin{equation}
 \frac{\partial U^{\Lam}}{\partial\Lam} = 
 \frac{(U^{\Lam})^2}{2\pi \sin^2 k_F} 
 \sum_{\om = \pm\Lam} \int_0^{2\pi} \frac{dp}{2\pi} \, f(p,\om) 
\end{equation}
where
\begin{equation}
 f(p,\om) = 
 \frac{2 \sin^2 k_F \, \sin^2 p}{(i\om - \xi^0_p)(-i\om - \xi^0_{-p})} -
 \frac{(\cos k_F - \cos p)^2}{(i\om - \xi^0_p)^2} +
 \frac{[\cos(2k_F) - \cos p]^2}
  {(i\om - \xi^0_{p-k_F})(i\om - \xi^0_{p+k_F})} \; .
\end{equation}
Here $\xi^0_k = -2 \cos k - \mu_0$ with $\mu_0 = - 2 \cos k_F$ is
the bare dispersion relation relative to the bare Fermi level.
Since $f(p,\om)$ can be written as a rational function of $\cos p$
and $\sin p$, the $p$-integral can be carried out analytically using
the substitution $z = e^{ip}$ and the residue theorem.
After a lengthy but straightforward calculation one obtains the
following result for the coefficient $h(\Lam)$ in (28):
\begin{eqnarray}
 h(\Lam) &=& - \frac{1}{2\pi} -
 \, \Re \bigg[ \frac{i}{2} \, (\mu_0 + i\Lam) 
 \, \sqrt{1 - \frac{4}{(\mu_0 + i\Lam)^2}} \nonumber \\[2mm]
 && \times \; \frac
 {3i\mu_0^4 - 10\mu_0^3\Lam - 12i\mu_0^2(\Lam^2 + 1) + 6\Lam^3\mu_0
  + 18\Lam\mu_0 + 6i\Lam^2 + i\Lam^4}
 {\pi (2\mu_0 + i\Lam)(4 - \mu_0^2 + \Lam^2 - 2i\Lam\mu_0)^2} \,
  \bigg] \; .
\end{eqnarray}
At half-filling ($\mu_0 = 0$), this reduces to
\begin{equation}
 h(\Lam) = - \frac{1}{2\pi} \, \left[ 1 -
 \Lam \, \frac{6 + \Lam^2}{(4 + \Lam^2)^{3/2}} \right] \; .
\end{equation}

\section{Calculation of $K_{\rho}$}

In this appendix we show how the Luttinger liquid parameter $K_{\rho}$,
which determines the critical exponents of Luttinger liquids, 
can be computed from the fixed point couplings as obtained from
the RG. A comparison of the RG result for $K_{\rho}$ with the
exact Bethe ansatz result for the bulk model (without impurity)
serves also as a check for the accuracy of our vertex renormalization. 
A relation between the fixed point couplings and $K_{\rho}$ can be
established via the exact solution of the fixed point Hamiltonian 
of Luttinger liquids, the Luttinger model.

For spinless fermions, $K_{\rho}$ is determined by the Luttinger model 
parameters $g$ and $v_F$ as
\begin{equation}
 K_{\rho} = \sqrt{\frac{1 - g/(2\pi v_F)}{1 + g/(2\pi v_F)}}
\end{equation}
where $g$ is the interaction between left and right movers and
$v_F$ the effective Fermi velocity of the model, that is the
slope of the (linear) dispersion relation, with a possible
shift due to interactions between particles moving in the same
direction ($g_4$-coupling) already included.\cite{Voi}
We therefore need to extract $g$ and $v_F$ from the RG flow in
the limit $\Lam \to 0$. In order to obtain $K_{\rho}$ correctly to
order $U^2$, it is sufficient to obtain $v_F$ correctly to linear
order in $U$.

The Luttinger model interaction $g$ and the fixed point coupling
$g^* = \Gam^{\Lam \to 0}_{k_F,-k_F;k_F,-k_F}$ from the RG 
are not simply identical, in contrast to what one might naively 
expect.
To find the true relation between $g$ and $g^*$, one has to take
into account that the forward scattering limit of the dynamical
2-particle vertex is generally not unique (in the absence of cutoffs),
and depends on whether momentum or frequency transfers tend to
zero first. This ambiguity is well-known in Fermi liquid theory,
where it leads to the distinction between quasi-particle interactions
and scattering amplitudes,\cite{NO} but is equally present in 
Luttinger liquids, for the same reason in all cases: the ambiguity
of the small momentum, small frequency limit of particle-hole
propagators contributing to the vertex function.
In the {\em dynamical limit}, where the momentum transfer 
$q$ vanishes first, the singular particle-hole propagators do not 
contribute. In Fermi liquids this limit yields the quasi-particle
interaction. 
In the opposite {\em static limit} the frequency transfer 
$\nu$ vanishes first and particle-hole propagators yield a finite 
contribution.   
In the presence of an infrared cutoff $\Lam > 0$ the forward
scattering limit of the vertex function is unique, since the
ambiguity in the particle-hole propagator is due to the infrared
pole of the single-particle propagator. 
Hence $\Gam^{\Lam}_{k_F,-k_F;k_F,-k_F}$ is well defined.
However, $\Gam^{\Lam}_{k_F,-k_F;k_F,-k_F}$ and also its limit 
for $\Lam \to 0$ depend on the choice for the cutoff function.
For a momentum cutoff, which excludes states with excitation energies
below $\Lam$ around the Fermi points, particle-hole excitations with
small momentum transfers $q$ are impossible. Hence particle-hole 
propagators with infinitesimal $q$ do not contribute to the vertex
at any $\Lam > 0$, such that $\Gam^{\Lam}_{k_F,-k_F;k_F,-k_F}$
converges to the dynamical forward scattering limit, which is simply
given by the bare coupling constant $g$ in the Luttinger model.
For a frequency cutoff the particle-hole propagators with vanishing
momentum and frequency transfer yield a finite contribution at
$\Lam > 0$, which tends to the static limit for $\Lam \to 0$. 
This can be seen directly by integrating 
${\sum_{\om = \pm\Lam} \int dp} \, [G^0_p(i\om)]^2$ 
over $\Lam$ from infinity to zero. 
Hence the vertex $\Gam^{\Lam}_{k_F,-k_F;k_F,-k_F}$
obtained from our frequency cutoff RG tends to the {\em static} 
forward scattering limit.

For the Luttinger model, the static forward scattering limit of the
vertex can be obtained from the dynamical effective interaction between
left and right movers $D(q,i\nu)$, which is defined as the sum of 
particle-hole chains
\begin{equation}
 D(q,i\nu) = 
 g + g \, \Pi^0_-(q,i\nu) \, g \, \Pi^0_+(q,i\nu) \, g + \dots =
 \frac{g}{1 - g^2 \, \Pi^0_-(q,i\nu) \, \Pi^0_+(q,i\nu)}
\end{equation}
where
\begin{equation}
 \Pi^0_{\pm}(q,i\nu) = 
 \pm \frac{1}{2\pi} \, \frac{q}{i\nu \mp v_F q}
\end{equation}
is the bare particle-hole bubble for right (+) and left (-) movers.
Note that only odd powers of $g$ contribute to the effective 
interaction between left and right movers.
This effective interaction appears naturally in the exact
solution of the Luttinger model via Ward identities.\cite{DL}
For the static limit one obtains
\begin{equation}
 \lim_{q \to 0} D(q,0) = \frac{g}{1 - [g/(2\pi v_F)]^2}
\end{equation}
which we identify with our fixed point coupling $g^*$ as obtained
from the RG with frequency cutoff. 
Inverting this relation between $g$ and $g^*$ we obtain
\begin{equation}
 g = \frac{2\pi v_F}{g^*} \, \left[ - \pi v_F +
 \sqrt{(\pi v_F)^2 + (g^*)^2} \right]
\end{equation}
For spinless fermions the difference between $g$ and $g^*$ appears 
only at third order in the coupling, but for models with spin the 
distinction becomes important already at second order.

The Fermi velocity $v_F$ can be computed from the (frequency
independent) self-energy in momentum space as
\begin{equation}
 v_F = v^0_F + \left. \partial_k \Sg_k \right|_{k_F} \; ,
\end{equation}
where $v^0_F = \partial_k \eps_k |_{k_F}$ is the bare Fermi
velocity. 
The self-energy is computed from the flow equation (32), which
can be rewritten in momentum space as
\begin{equation}
 \frac{\partial}{\partial\Lam} \, \Sg^{\Lam}_k =
 - \frac{U^{\Lam}}{\pi} \sum_{\om = \pm\Lam} \int \frac{dp}{2\pi} \, 
 \frac{1 - \cos(k-p)}{i\om - \xi_p - \Sg^{\Lam}_p}
\end{equation}
where $\xi_p = \eps_p - \mu$. The chemical potential $\mu$ has to be 
fixed by the final condition $\xi_{k_F} + \Sg_{k_F} = 0$, where 
$k_F = \pi n$ depends only on the density, not the interaction.
From the tridiagonal structure of $\Sg$ in real space, but also from 
the above expression it follows that $\Sg^{\Lam}_k$ has the form 
$\Sg^{\Lam}_k = a^{\Lam} + b^{\Lam} \cos k$.
The functional flow equation for $\Sg^{\Lam}_k$ yields a coupled 
set of ordinary flow equations for the coefficients $a^{\Lam}$ and 
$b^{\Lam}$, with initial conditions $a^{\Lam_0} = U$ and 
$b^{\Lam_0} = 0$. 
The momentum integrals can be evaluated analytically
via the residue theorem, such that the remaining set of two coupled
differential equations (with $U^{\Lam}$ as input) can be easily 
solved numerically.
The result for $v_F$ is correct at least to first order in $U$, but
not necessarily to second order, since our simplified parametrization
captures the 2-particle vertex correctly to second order only at the 
Fermi points.

Inserting $g$ and $v_F$ into the Luttinger model formula (B1) we
can now compute $K_{\rho}$ as a function of $U$ and density for the 
microscopic spinless fermion model. 
In Fig.\ 16 we show results for $K_{\rho}(U)$ for various fixed 
densities, which are compared to the exact result from the Bethe 
ansatz solution \cite{Hal} in the inset. 
The RG results are indeed correct to order $U^2$ for small $U$, as 
expected, and they are surprisingly accurate also for larger values 
of $U$.

\section{Numerical solution of ${\bf\Sg}$ flow}

For $\Lam < \Lam_0 < \infty$ the flow equation for the self-energy
(16) can be written as
\begin{equation} \label{eq:appc:sigmareal}
  \frac{\partial}{\partial\Lam} \Sg_{1',1}^\Lam =
  - \frac{1}{2\pi} \sum_{2,2'} \Gam_{1',2';1,2}^\Lam 
  \cdot 2 \, \Re \left[
    \tilde G_{2,2'}^\Lam(i\Lam) \right] \; .
\end{equation}
In order to compute its right-hand side, one needs to invert the 
tridiagonal matrix
\begin{equation} \label{eq:appc:tridiag}
 T = G_0^{-1}(i\Lam) - \Sg^\Lam \; ,
\end{equation}
where $T$ is complex symmetric (\emph{not} hermitean) with diagonal 
elements $a_i := i\Lam + \mu - \Sg_{i,i}^\Lam$, $i=1,\ldots,L$, and 
first off-diagonal elements $b_i := t - \Sg_{i,i+1}^\Lam$, 
$i=1,\ldots,L-1$.  
Note that $\Im(a_i) = \Lam > 0$ such that $T$ is non-singular 
and its inverse well-defined.

The inverse $\tilde G^\Lambda(i\Lam) = T^{-1}$ is not tridiagonal
but a full matrix which can be computed by standard methods in
$\mathcal O(L^2)$ time. However, for an interaction that does
not extend beyond nearest neighbors on the lattice, only the 
tridiagonal part of $\tilde G$ is required, which can be computed 
in $\mathcal O(L)$ time, such that much larger lattices can be
treated.  
We shall first explain how this is done and then present the 
resulting algorithm that can directly be incorporated into a
computer program.

Under certain assumptions (see below), a matrix can be uniquely
factorized into a lower unit triangular matrix $L$, a diagonal matrix
$D$, and an upper unit triangular matrix $U$ (``LDU factorisation''):
$T=L\cdot D\cdot U$.\cite{Num}  
For a tridiagonal matrix $T$ the unit triangular matrices $L$
and $U$ are in fact unit bidiagonal: their matrix elements are one 
on the diagonal, and only the first off-diagonal is nonzero.  
Since our $T$ is symmetric we have $L=U^T$.
Thus we obtain a factorization of the form 
\begin{equation*}
  T = U^{+T} D^+ U^+
  = \begin{pmatrix}
    1 & & & \\
    U_1^+ & 1 & & \\
    & U_2^+ & 1 & \\
    & & \ddots & \ddots
  \end{pmatrix}
  \begin{pmatrix}
    D_1^+ & & & \\
    & D_2^+ & & \\
    & & D_3^+ & \\
    & & & \ddots
  \end{pmatrix}
  \begin{pmatrix}
    1 & U_1^+ & & \\
    & 1 & U_2^+ & \\
    & & 1 & \ddots \\
    & & & \ddots
  \end{pmatrix}
\end{equation*}
where the label ``$+$'' distinguishes this factorization from 
another one used below.
The prescription to compute the elements $D_i^+$ and $U_i^+$ is well
known and can be found for example in Ref.\ \cite{Num}. 
Starting in the upper left corner one proceeds to increasing row and 
column numbers until one arrives at the lower right corner of $T$:
\begin{equation} \label{eq:appc:ldu}
  D^+_1:=a_1, \quad
  U^+_i:=b_i/D^+_i, \quad
  D^+_{i+1}:=a_{i+1}-b_i U^+_i \quad
  (i=1,\ldots,L-1) \; .
\end{equation}
This works well since in our case $\Im(D^+_i) \geq \Lam > 0$, such
that one never divides by zero.

To compute the inverse $\tilde G=T^{-1}$, one could directly 
calculate $(U^+)^{-1} (D^+)^{-1} (U^{+T})^{-1}$.  
It is however easier and more accurate to find the inverse by solving 
the linear system of equations $T \tilde G = 1$, where $1$ is the 
identity matrix, by ``backsubstitution''. To be specific,
consider the $i^{\mathrm{th}}$ column vector $\tilde G_{\cdot,i}$ of
$\tilde G$:
\begin{equation} \label{eq:appc:backsubst}
  e_i = T \tilde G_{\cdot,i} = U^{+T} (D^+ U^+ \tilde G_{\cdot,i})
  = U^{+T} g_i,\quad
  U^+ \tilde G_{\cdot,i} = (D^+)^{-1} g_i
\end{equation}
where $e_i$ is the $i^{\mathrm{th}}$ unit vector.  The first step is
to solve the linear system $U^{+T} g_i = e_i$ for $g_i$, and the
second step to solve $U^+ \tilde G_{\cdot,i} = (D^+)^{-1} g_i$ for
$\tilde G_{\cdot,i}$.  To solve a tridiagonal linear system for one
vector takes $\mathcal O(L)$ time, so solving for the full inverse
matrix $\tilde G$ takes $\mathcal O(L^2)$ time.

Now we shall derive an algorithm to compute the elements of $g_i$ and
$\tilde G_{\cdot,i}$.  Begin with the last column $i=L$: $U^{+T}
g_L=e_L$ can be solved from the first to the last row and gives
$g_L=e_L$.  Next $U^+ \tilde G_{\cdot,L} = (D^+)^{-1} e_L$ can be
solved starting from the last row, $\tilde G_{L,L} = 1/D_L^+$.  From
there one can work upwards by backsubstitution, $\tilde G_{j,L} =
-U_j^+ \tilde G_{j+1,L}$ ($j=1,\ldots,L-1$).  For the other columns
$i<L$, one cannot take the shortcut and has to solve both linear
systems for $g_i$ and $\tilde G_{\cdot,i}$.  But it is now important
to realize that for any column vector $\tilde G_{\cdot,i+1}$, if we
somehow know the diagonal element $\tilde G_{i+1,i+1}$, the next
element above the diagonal is
\begin{equation} \label{eq:appc:fromdiag}
  \tilde G_{i,i+1} = -U_i^+ \tilde G_{i+1,i+1} \quad 
  (i=1,\ldots,L-1) \; .
\end{equation}
Thus we have a prescription how to go up one row in $\tilde G$.
Together with the symmetry of $\tilde G$, that is
$\tilde G_{i,i+1} = \tilde G_{i+1,i}$, which follows from the 
symmetry of $T$, we get the first off-diagonal element one column 
to the left \emph{without} solving the two linear systems in
(\ref{eq:appc:backsubst}). So it is possible to compute directly the
tridiagonal part of the inverse.  However, there is another algorithm
which is much more accurate for near-singular matrices at the end of
the RG flow: the double factorization.\cite{Meu}
It does not rely on the symmetry of $\tilde G$ but uses the 
complementary ``UDL'' factorization
\begin{equation} \label{eq:appc:udlfactor}
  T = U^- D^- L^- = U^- D^- U^{-T} \; ,
\end{equation}
where the matrix elements are obtained as
\begin{equation} \label{eq:appc:udl}
  D^-_L:=a_L, \quad
  U^-_i:=b_i/D^-_{i+1}, \quad
  D^-_i:=a_i-b_i U^-_i \quad
  (i=L-1,\ldots,1) \; .
\end{equation}
We proceed as for the LDU factorization above and get
\begin{eqnarray}
  \label{eq:appc:backsubst0}
  \tilde G_{1,1}&=&1/D^-_1 \\
  \label{eq:appc:todiag}
  \tilde G_{i,i+1}&=&-U^-_i\cdot\tilde G_{i,i} \; .
\end{eqnarray}
We can combine equations (\ref{eq:appc:fromdiag}) and
(\ref{eq:appc:todiag}) to relate consecutive diagonal elements:
\begin{equation} \label{eq:appc:backsubst2}
  \tilde G_{i+1,i+1} = -\tilde G_{i,i+1}/U^+_i =
  \tilde G_{i,i}\cdot U^-_i/U^+_i =
  \tilde G_{i,i}\cdot D^+_i/D^-_{i+1} \; .
\end{equation}
Thus we start with (\ref{eq:appc:backsubst0})  and
use the $U^-D^-L^-$ decomposition to go one matrix element to
the right in the inverse matrix, from the diagonal to the first
off-diagonal (\ref{eq:appc:todiag}), while the $L^+D^+U^+$
decomposition allows to go down by one, back to the next diagonal
element (\ref{eq:appc:fromdiag}). There is no need to compute
the full inverse matrix.  

One can implement the algorithm without knowing the
derivation by using equations (\ref{eq:appc:ldu}) and
(\ref{eq:appc:udl}) - (\ref{eq:appc:backsubst2}). One can
further eliminate the $U$'s using equations (\ref{eq:appc:ldu}) and
(\ref{eq:appc:udl}) and implement the algorithm such that only the 
input vectors $a_i$, $b_i$ and the output vectors $\tilde G_{i,i}$, 
$\tilde G_{i,i+1}$ enter the temporary storage.  
This double factorization is numerically accurate to more than 10 
significant digits (using double precision) even for large lattices 
($10^6$ sites) and almost singular matrices with $|a_i| \sim 
10^{-15}$ which appear at the end of the flow for half filling.

The right hand side of the flow equations (41) for the density 
response vertex $R^{\Lam}$ can be computed in ${\cal O}(L)$ time
using the fact that the upper triangular part of the inverse of
a tridiagonal matrix (here $\tG$) is the upper triangular part
of the outer product of two vectors.\cite{Meu,Ens}

\end{appendix}


\vfill\eject


\begin{figure}
\center
\vskip 1cm
\epsfig{file=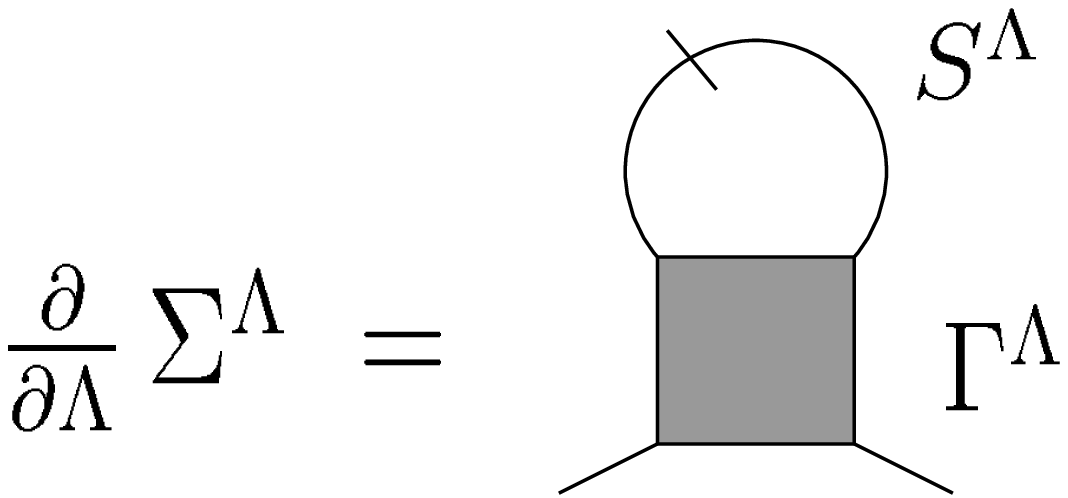,width=4cm}
\vskip 1cm 
\caption{Flow equation for the selfenergy $\Sg^{\Lam}$.}
\end{figure}

\begin{figure}
\center
\vskip 1cm
\epsfig{file=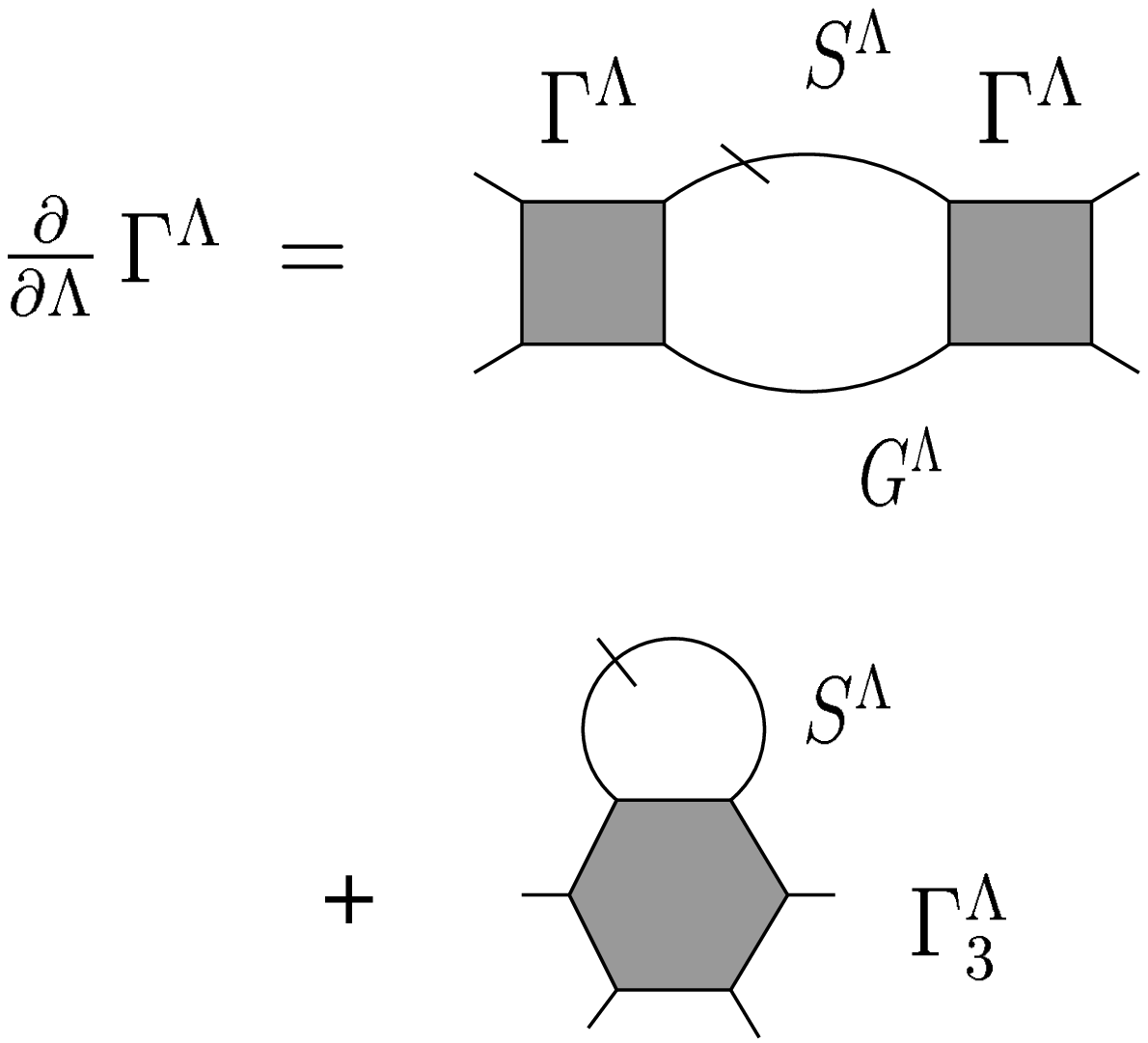,width=6cm}
\vskip 1cm 
\caption{Flow equation for the 2-particle vertex $\Gam^{\Lam}$.}
\end{figure}

\vfill\eject

\begin{figure}
\center
\epsfig{file=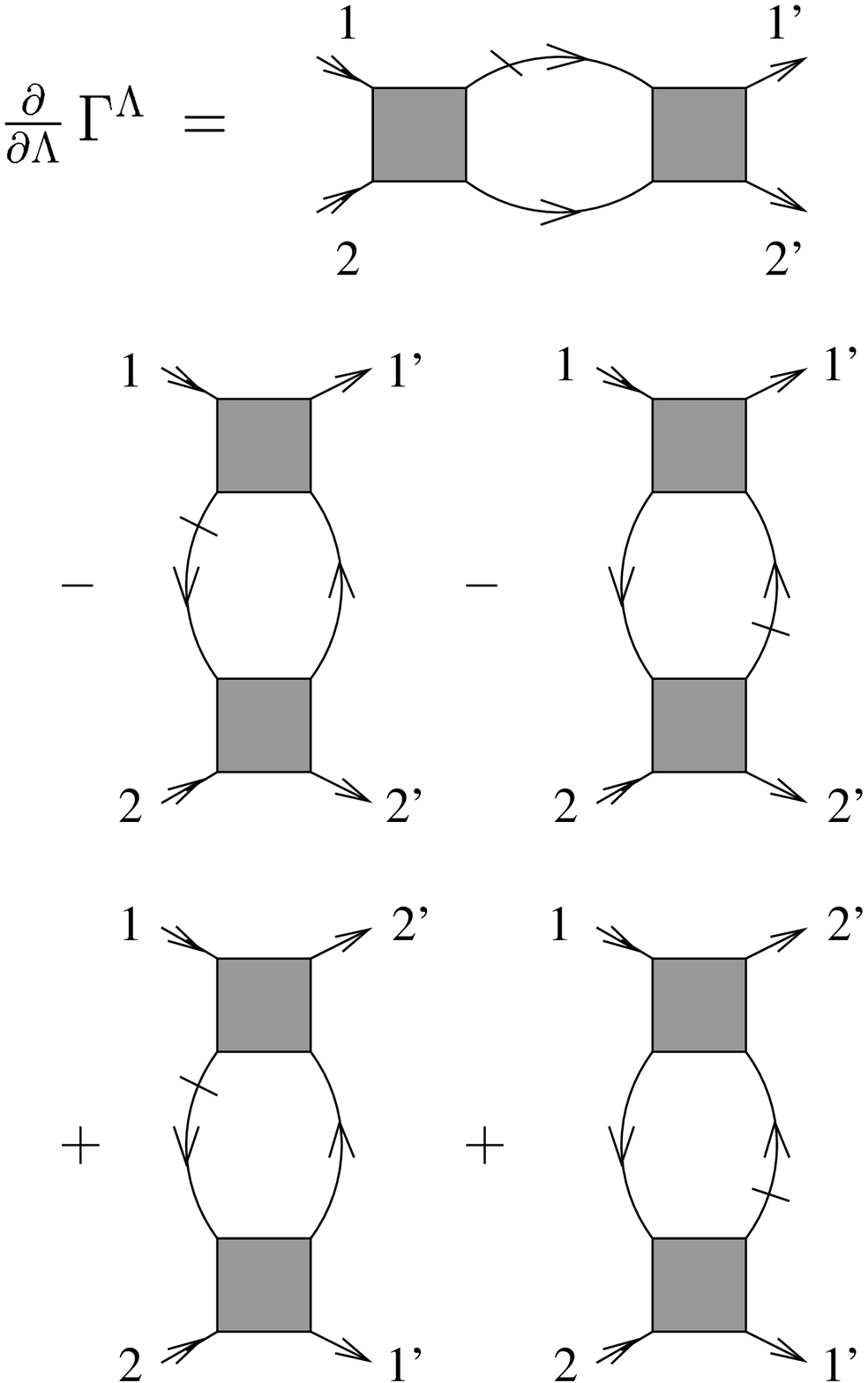,width=6cm}
\vskip 1cm 
\caption{Flow equation for the 2-particle vertex at 1-loop level
 with particle-particle and particle-hole channels written 
 explicitly.}
\end{figure}

\begin{figure}
\center
\epsfig{file=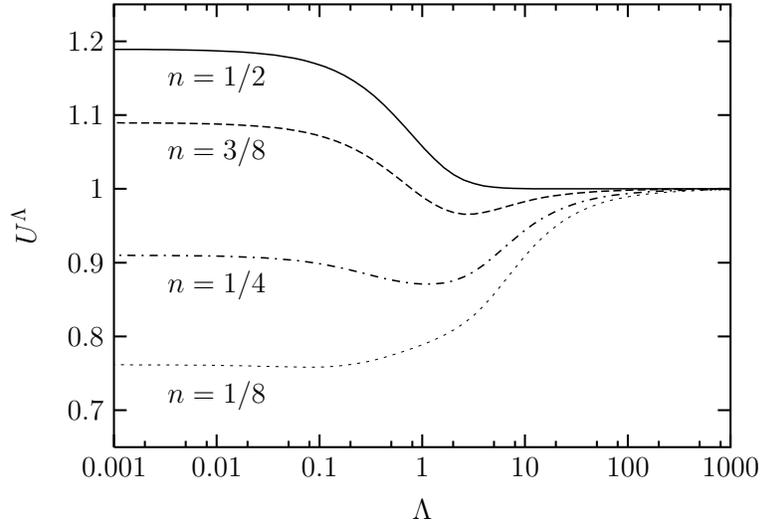,width=10cm}
\vskip 5mm
\caption{Flow of the renormalized nearest neighbor interaction
 $U^{\Lam}$ for the spinless fermion model, for $U=1$ and 
 various densities $n$.}
\end{figure}

\begin{figure}
\center
\epsfig{file=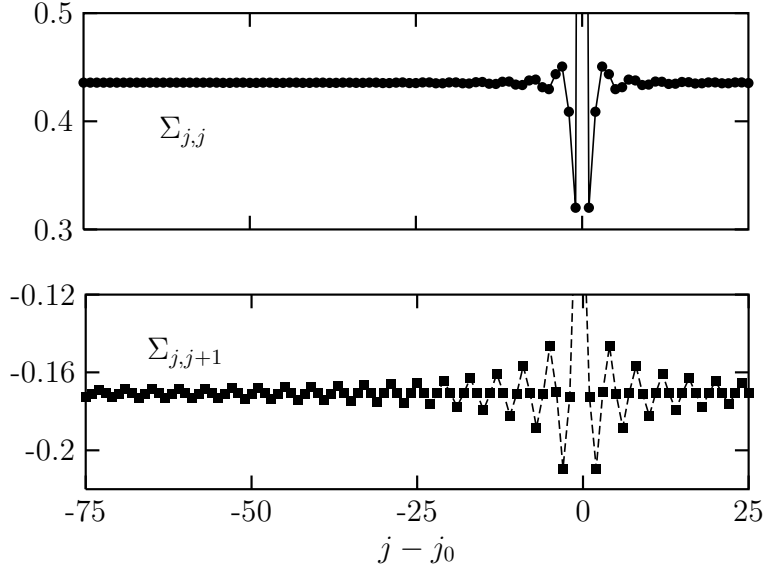,width=10cm}
\vskip 5mm
\caption{Self-energy near a site impurity of strength $V=1.5$
 for the spinless fermion model at quarter-filling and 
 interaction strength $U=1$; the impurity is situated at the
 center of a chain with $L = 1025$ sites.}
\end{figure}

\begin{figure}
\center
\epsfig{file=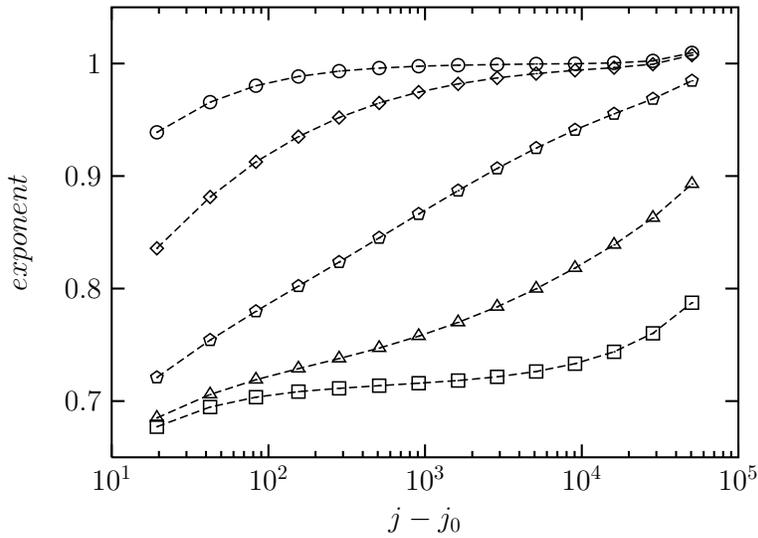,width=10cm}
\vskip 5mm
\caption{Effective exponent for the decay of oscillations of 
 $\Sg_{j,j}$ as a function of the distance from a site impurity 
 of strengths $V=0.01$, $0.1$, $0.3$, $1$, $10$ (from bottom 
 to top), for the spinless fermion model at half-filling and 
 interaction strength $U=1$; 
 the impurity is situated at the center of a chain with 
 $L=2^{18}+1$ sites.}
\end{figure}

\begin{figure}
\center
\epsfig{file=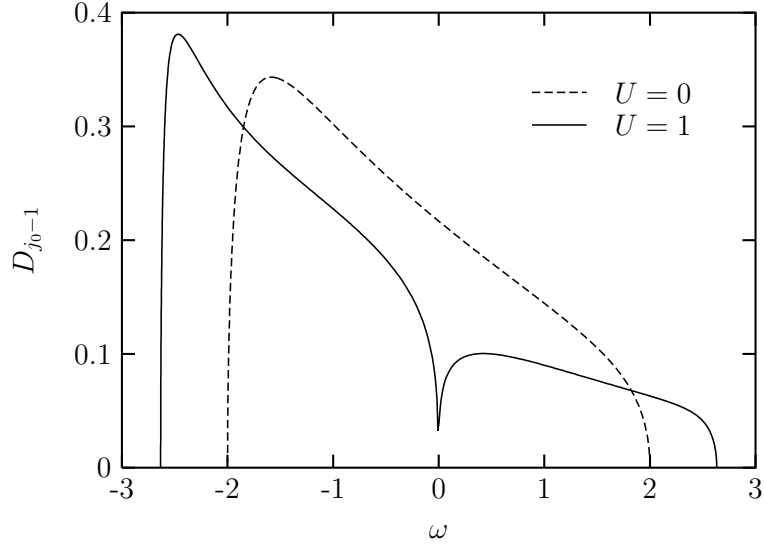,width=10cm}
\vskip 5mm
\caption{Local density of states on the site next to a site
 impurity of strength $V=1.5$ for spinless fermions at 
 half-filling and $U=1$; the impurity is situated at the
 center of a chain with $L = 1025$ sites; the non-interacting
 case $U=0$ is shown for comparison.}
\end{figure}

\begin{figure}
\center
\epsfig{file=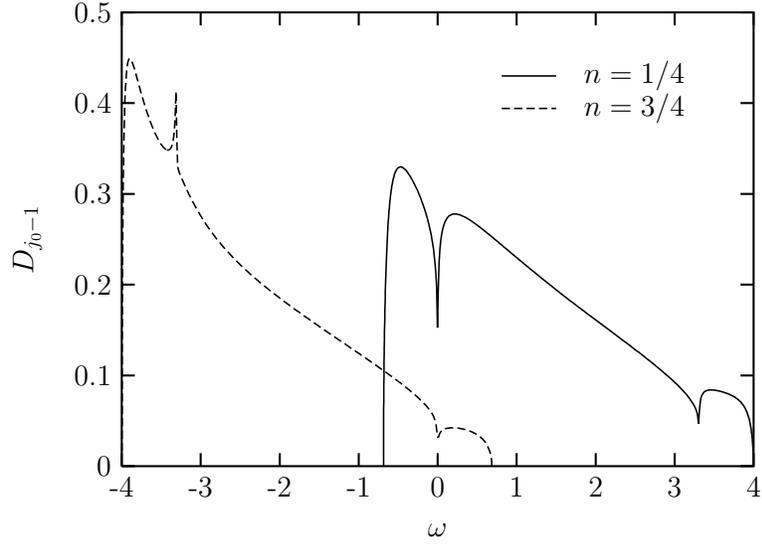,width=10cm}
\vskip 5mm
\caption{Local density of states on the site next to a site
 impurity as in Fig.\ 7 (same parameters), but now for 
 densities $n=1/4$ and $n=3/4$.}
\end{figure}

\begin{figure}
\center
\epsfig{file=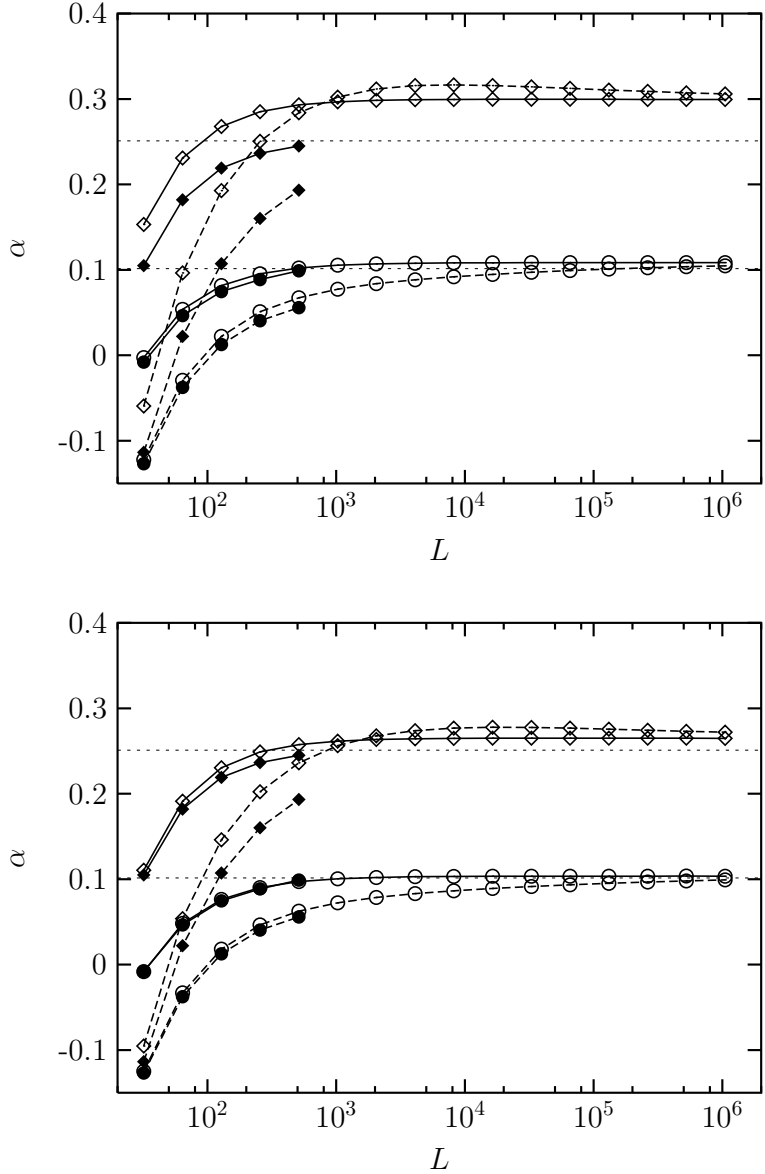,width=10cm}
\vskip 5mm
\caption{Logarithmic derivative of the spectral weight at the Fermi 
 level near a boundary (solid lines) or hopping impurity (dashed
 lines) as a function of system size $L$, for spinless fermions at 
 quarter-filling and interaction strength $U=0.5$ (circles) or 
 $U=1.5$ (squares); 
 \emph{upper panel}: without vertex renormalization,
 \emph{lower panel}: with vertex renormalization;
 the open symbols are fRG, the filled symbols DMRG 
 results;
 the horizontal lines represent the exact boundary exponents for
 $U=0.5$ and $U=1.5$. 
 In the boundary case (solid lines) the spectral weight has been 
 taken on the first site of a homogeneous chain,
 in the impurity case (dashed lines) on one of the two sites 
 next to a hopping impurity $t'=0.5$ in the center of the chain.}
\end{figure}

\begin{figure}
\center
\epsfig{file=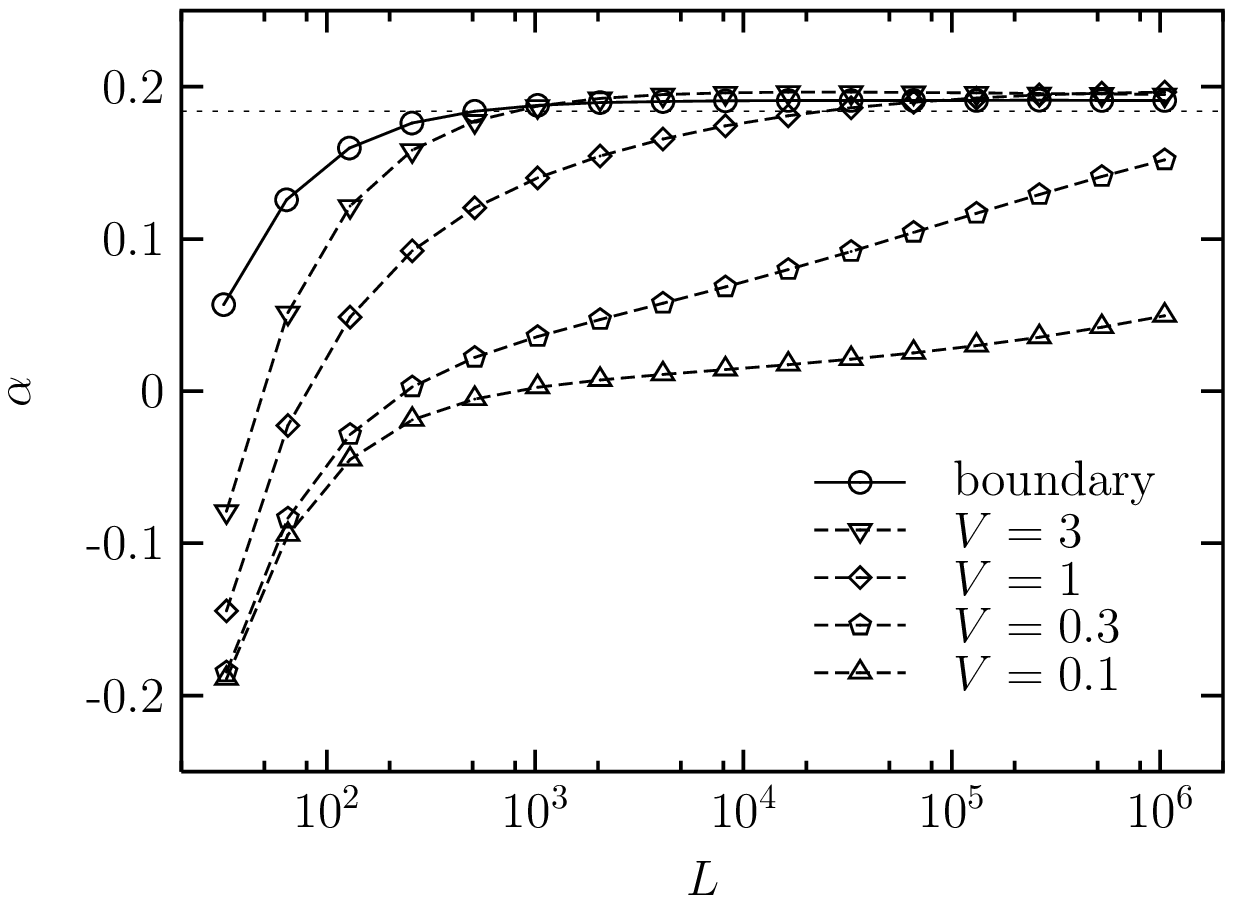,width=10cm}
\vskip 5mm
\caption{Logarithmic derivative of the spectral weight at the 
 Fermi level near a boundary (solid line) or near a site impurity 
 (dashed lines) as a function of system size $L$, for spinless 
 fermions at quarter-filling and interaction strength $U=1$.
 In the boundary case the spectral weight has been 
 taken on the first site of a homogeneous chain, in the impurity 
 case on the site next to a site impurity of strength $V$ in the 
 center of the chain. 
 The horizontal line represents the exact boundary exponent for
 $U=1$.}
\end{figure}

\begin{figure}
\center
\epsfig{file=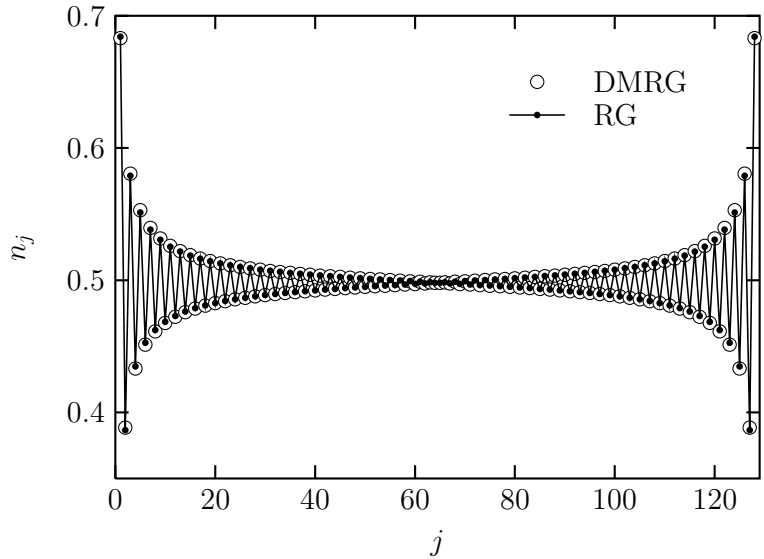,width=10cm}
\vskip 5mm
\caption{Density profile $n_j$ for a spinless fermion chain with
 $128$ sites and interaction strength $U=1$ at half-filling;
 fRG results are compared to exact DMRG results.}

\end{figure}

\begin{figure}
\center
\epsfig{file=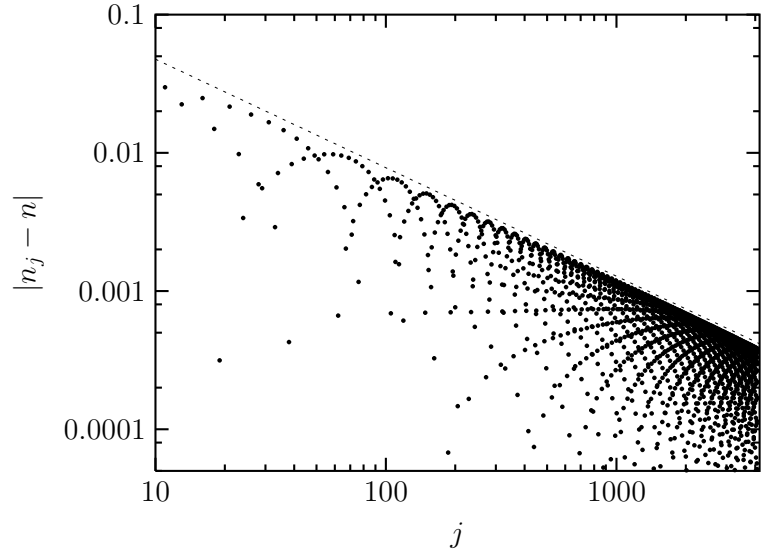,width=10cm}
\vskip 5mm
\caption{Density modulation $|n_j - n|$ as a function of the 
 distance from a boundary,
 for spinless fermions with interaction strength $U=1$ and 
 (average) density $n=0.393$ on a chain with $8192$ sites;
 the broken straight line is a power-law fit to the envelope
 of the oscillation amplitudes.}
\end{figure}

\begin{figure}
\center
\epsfig{file=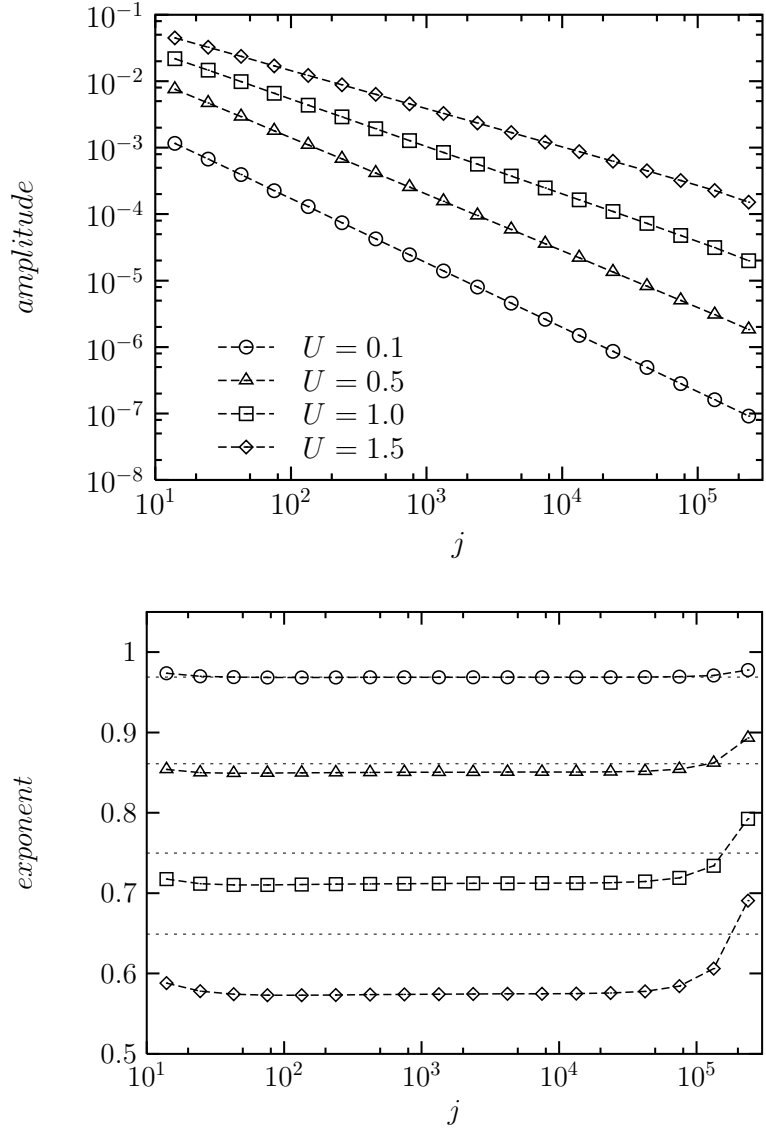,width=10cm}
\vskip 5mm
\caption{Amplitude (envelope) of oscillations of the density 
 profile $n_j$ induced by a boundary as a function of the distance 
 from the boundary, for spinless fermions with various interaction
 strengths $U$ at half-filling; the interacting chain with
 $2^{19}+1$ sites is coupled to a semi-infinite non-interacting
 lead at one end (opposite to the boundary); 
 \emph{upper panel}: log-log plot of the amplitude,
 \emph{lower panel}: effective exponents for the decay, and the 
 exact asymptotic exponents as horizontal lines.}
\end{figure}

\begin{figure}
\center
\epsfig{file=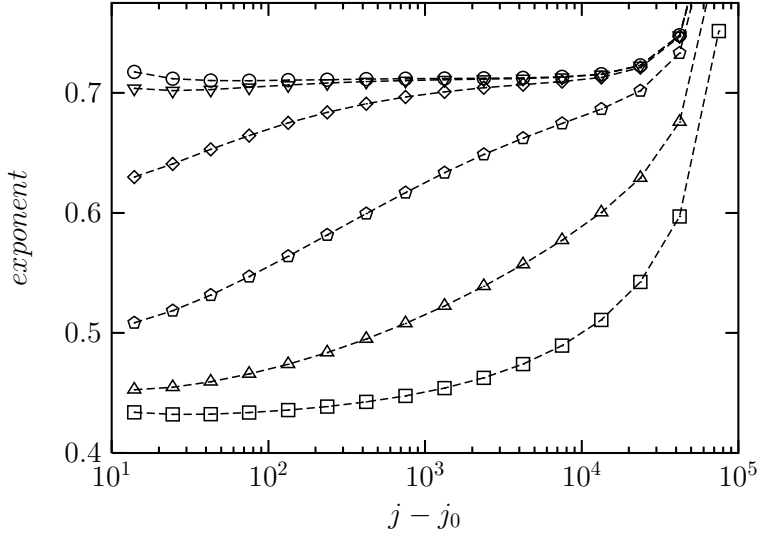,width=10cm}
\vskip 5mm
\caption{Effective exponent for the decay of density oscillations
 as a function of the distance from a site impurity of strengths
 $V=0.01$, $0.1$, $0.3$, $1$, $3$, $10$ (from bottom to top); 
 the impurity is situated at the center of a spinless 
 fermion chain with $2^{18}+1$ sites and interaction strength 
 $U=1$ at half-filling; the interacting chain is coupled to
 semi-infinite non-interacting leads at both ends.}
\end{figure}

\begin{figure}
\center
\epsfig{file=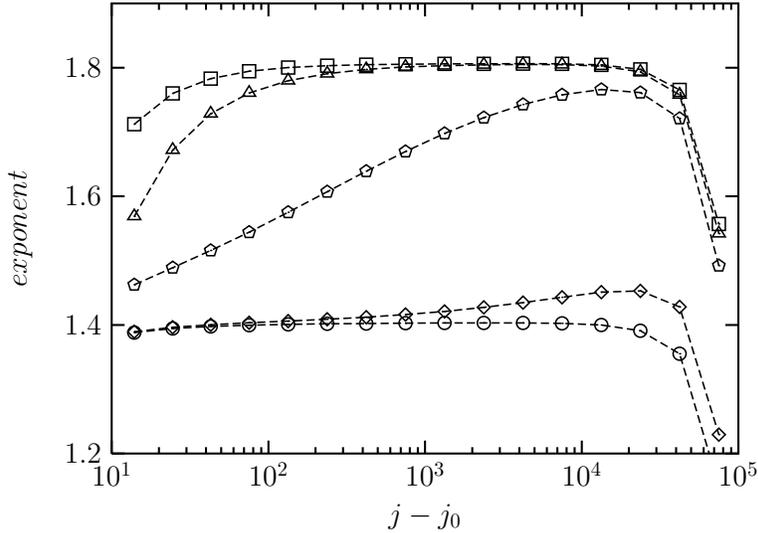,width=10cm}
\vskip 5mm
\caption{Effective exponent for the decay of density oscillations
 as a function of the distance from a site impurity of strengths
 $V=0.1$, $1$, $10$, $100$, $1000$ (from top to bottom) for the 
 same chain as in Fig.\ 14 but now with an \emph{attractive} 
 interaction, $U=-1$.}
\end{figure}

\begin{figure}
\center
\epsfig{file=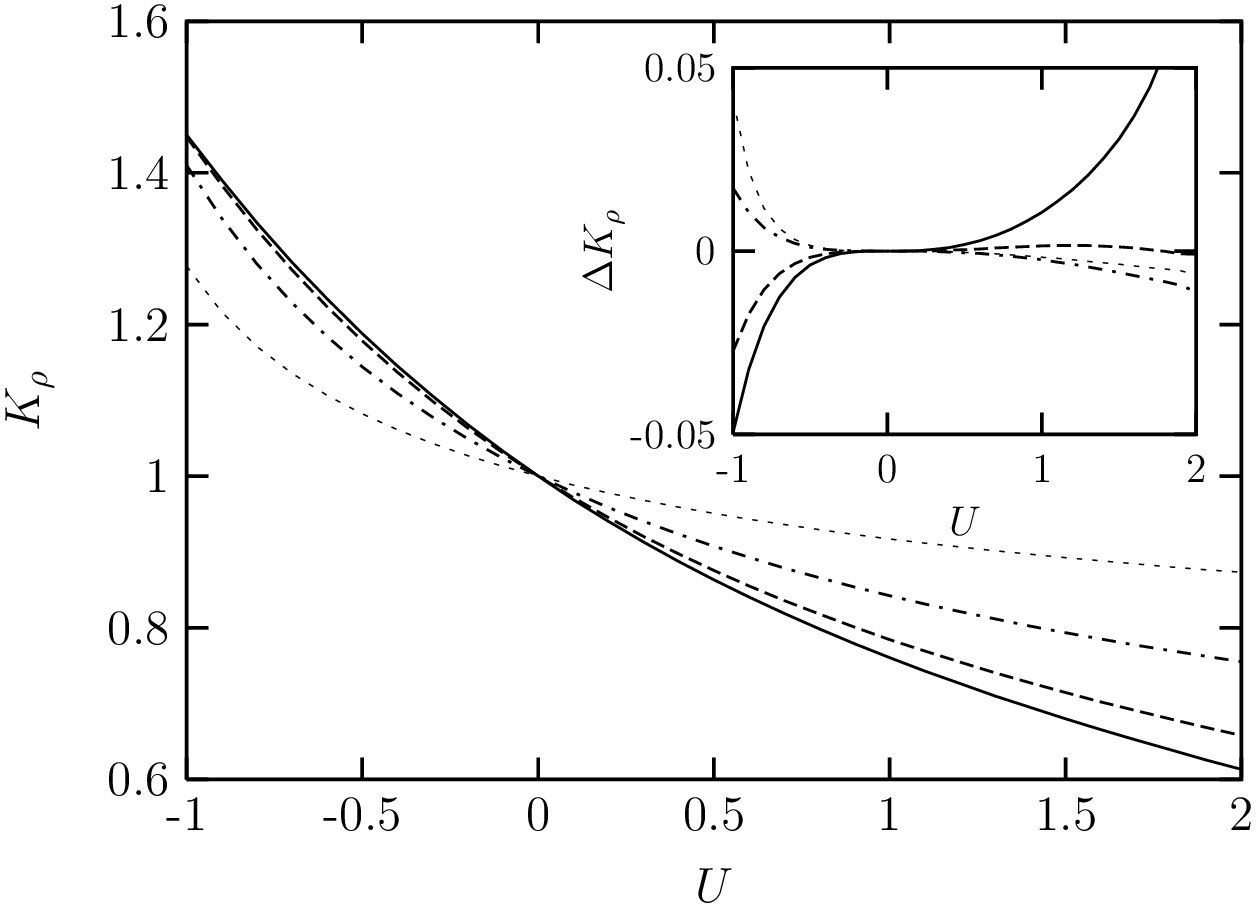,width=10cm}
\vskip 5mm 
\caption{Luttinger liquid parameter $K_{\rho}$ as a function of $U$
 at various densities (as in Fig.\ 4) for the spinless fermion model; 
 the inset shows the difference between the RG result and the 
 exact Bethe ansatz result for $K_{\rho}$.}
\end{figure}


\begin{thebibliography}{99}
\bibitem{fn1} Vertex renormalization has already been applied in 
 two letters on transport properties, Ref.\ [20,21], but without 
 any derivation of the method.
\end{thebibliography}
\end{document}